\documentclass{JHEP3} 

\usepackage{epsfig,multicol}
\usepackage{amsmath, amssymb}
\usepackage{concmath, palatino}

\newcommand\fverb{\setbox\pippobox=\hbox\bgroup\verb}
\newcommand\fverbdo{\egroup\medskip\noindent%
			\fbox{\unhbox\pippobox}\ }
\newcommand\fverbit{\egroup\item[\fbox{\unhbox\pippobox}]}
\newbox\pippobox

\newcommand{\be}{\begin{equation}}
\newcommand{\ee}{\end{equation}}
\newcommand{\ba}{\begin{eqnarray}}
\newcommand{\ea}{\end{eqnarray}}

\newcommand{\pint}{\makebox[0pt][l]{\hspace{2.4pt}$-$}\int}

\newcommand{\refeq}[1]{Eq.~(\ref{eq:#1})}

\newcommand{\la}{\longrightarrow}

\newcommand{\ads}{AdS_5\times S^5}

\title{On the strong coupling expansion in the $\mathfrak{su}(1|1)$ sector of ${\cal N}=4$ SYM}

\author{Matteo Beccaria\\
  Dipartimento di Fisica, Universita' del Salento,
  Via Arnesano, 73100 Lecce\\
  INFN, Sezione di Lecce\\
  E-mail: \email{matteo.beccaria@le.infn.it}}

\abstract{
We consider the anomalous dimension of the fermionic highest states $\mbox{Tr}\,\psi^L$ in the 
$\mathfrak{su}(1|1)$ sector of ${\cal N}=4$ SYM at strong coupling. In the thermodynamical $L\to\infty$ limit it is described 
by a BES-like integral equation recently proposed by Rej, Staudacher and Zieme. 
The strong coupling regime of this equation is analyzed numerically and analytically by Neumann expansion methods which have been developed for the 
$\mathfrak{sl}(2)$ sector. We compute analytically the first two terms of the strong coupling expansion and present numerical results
for the next correction. We illustrate various specific features valid for the $\mathfrak{su}(1|1)$ sector. In particular, at next-to-leading 
order, we find and solve a singular integral equation describing the scaling continuum limit of the Neumann coefficients.
}

\begin{document} 

\section{Introduction}

The planar maximally supersymmetric ${\cal N}=4$ SYM theory has an intriguing integrability governing the 
renormalization mixing of its composite operators~\cite{Minahan:2002ve,Beisert:2003yb,Beisert:2003tq}.
The associated integrable Hamiltonian is the dilatation generator of  $\mathfrak{psu}(2,2|4)$ whose eigenvalues are the scaling dimensions. By AdS/CFT 
duality~\cite{Maldacena:1997re,Gubser:1998bc,Witten:1998qj}, its integrability properties are related to that of those of the dual type 
IIB superstring on $\ads$~\cite{Bena:2003wd}.

All order Bethe Ansatz equations connecting the two sides of the correspondence have been 
proposed in~\cite{Beisert:2006ez}. It is well known that these equations are asymptotic. The weak coupling perturbative expansion of anomalous dimensions 
is correct up to a certain wrapping order depending on the
length $L$ of the operator that one is considering. In the thermodynamical limit  $L\to\infty$ it is reasonable to assume that wrapping does not
occur and it is possible to derive integral equations~\cite{Beisert:2006ez,Rej:2007vm} describing exactly the flow of anomalous dimensions 
from weak to strong gauge coupling $g$ defined as 
\be
g^2 = \frac{\lambda}{16\,\pi^2},
\ee
where $\lambda$ is the 't Hooft planar coupling.

The Bethe equations describe  integrable elastic scattering of {\em internal} elementary excitations in terms of $2\to 2$ processes governed by 
a specific $S$-matrix~\cite{Staudacher:2004tk}. The $S$-matrix is partially determined by $\mathfrak{psu}(2,2|4)$ symmetry~\cite{Beisert:2005fw}
which completely fixes it up to a scalar factor, the dressing phase. It must obey the crossing relation of~\cite{Janik:2006dc}. 
The strong coupling expansion of the dressing phase can be matched to string theory calculations leading to the AFS
leading approximation~\cite{Arutyunov:2004vx}, and to the next-to-leading correction~\cite{Hernandez:2006tk}.
The crucial results of~\cite{Beisert:2006ib} allowed to obtain the current complete form of the dressing phase proposed by 
Beisert, Eden, and Staudacher (BES)~\cite{Beisert:2006ez}. 

The BES dressing factor is a clever analytical continuation of the known information 
on the string side of AdS/CFT down to the weak coupling regime. If correct, it permits in principle to extract the 
perturbative weak-strong coupling expansions on the two sides from the analysis of the Bethe Ansatz equations bypassing the painful high-order
standard diagrammatic expansions.

Recently, the BES formulation has been hardly tested looking at twist operators in the $\mathfrak{sl}(2)$ subsector of ${\cal N}=4$ SYM.
For these operators, one can consider the large spin $N$ behavior of the quantum anomalous dimension~\cite{Belitsky:2006en,Alday:2007mf}
\be
\Delta_N(g) = f(g)\,\log\,N + {\cal O}(N^0).
\ee
The so-called scaling function $f(g)$ is related to the cusp anomalous dimension introduced in~\cite{Korchemsky:1988si,Korchemsky:1992xv} and related to the 
universal soft gauge boson emission. It is known 
up to three loops in the gauge theory~\cite{cusp-gauge}. At four loops, it is known numerically with high precision~\cite{cusp-four}.
These results are in full agreement with the BES Bethe Ansatz calculation~\cite{Beisert:2006ez}. 
Remarkably, the weak coupling expansion of the BES integral equation can be extended to high orders with minor effort.

At strong coupling, the expansion of the scaling function can be performed on the string side and the calculations 
of~\cite{cusp-strong} leads to the expression 
\be
\label{eq:cusp-strong}
f(g) = 4\,g -\frac{3\,\log\,2}{\pi}-\frac{\rm K}{2\,\pi^2}\,\frac{1}{g} + \cdots~,
\ee
where $K$ is Catalan's constant.
We write the explicit form of this strong coupling expansion to emphasize that it appears to be an expansion in integer inverse powers of $g$.

Surprisingly, the analytic derivation of this expansion is rather non-trivial for reasons explained in~\cite{Kostov:2007kx}.
A direct numerical analysis of the BES equation has been proposed in~\cite{Benna:2006nd} and led to a
clear numerical approximation of the first three coefficient of the strong coupling expansion in agreement with \refeq{cusp-strong}. 
The analytical derivation of the first coefficient
has been performed in~\cite{Alday:2007qf,Kotikov:2006ts} starting from the BES equation. 
The analysis is greatly simplified if the BES dressing phase is replaced by its leading 
strong coupling term, the AFS phase~\cite{Beccaria:2007tk}. 
Honestly, this is not totally satisfactory since the main concern about the BES phase is 
its correct analytic continuation from weak to strong coupling and one would like to recover the AFS phase back from the BES phase.
A solution to this technical problem has been proposed in the paper~\cite{Kostov:2007kx} which also deals with the other rank-1 
$\mathfrak{su}(2)$ and $\mathfrak{su}(1|1)$ closed subsectors.

If the aim is that of deriving the expansion of \refeq{cusp-strong} by efficient integrability techniques, then these investigations 
definitely suggested to analyze the strong coupling expansion using the string Bethe equations leaving aside the BES equation. 
This approach was successfully adopted in ~\cite{Casteill:2007ct} where the next-to-leading coefficient in \refeq{cusp-strong} was computed. 
The same result has also been obtained in~\cite{Belitsky:2007kf} in the framework of the Baxter equation.

However, quite surprisingly, this intriguing story is now basically concluded by going back to the  BES formulation.
In the beautiful paper~\cite{Basso:2007wd}, Basso and Korchemsky showed that the full strong coupling expansion of the scaling function 
can be recovered by an improved analysis of the BES equation. The new ingredient are crucial scaling assumptions inspired by the 
numerical solution obtained by the methods of~\cite{Benna:2006nd}. Recently these assumptions have been put on rigorous footings in~\cite{Kostov:2008ax}.

Given this rather rich landscape of methods and results, we believe that it would be interesting to 
adopt the same viewpoint in the case of other sectors of ${\cal N}=4$ SYM. Here, we consider the $\mathfrak{su}(1|1)$ sector (see~\cite{Swanson} for 
its near pp-wave spectrum) and its maximally
excited states of the form 
\be
{\cal O}_L = \mbox{Tr}\, \psi^L,
\ee
where $\psi$ is a gaugino component. These operators are eigenstates of the dilatation operator for each length (odd) $L$.
The specific problem of deriving the strong coupling expansion of their anomalous dimensions $\Delta_L$ has been treated in some details
in~\cite{Arutyunov:2006av}. One motivation of this study was to understand if the Gubser-Klebanov-Polyakov $\Delta\sim \lambda^{1/4}$ law (GKP)~\cite{Gubser:2002tv}
is correct for these operators.
A numerical and analytical analysis of the string Bethe equations with AFS dressing~\cite{Beccaria:2006td} (confirmed 
by the later calculation of  \cite{Kostov:2007kx}) proved that the GKP law is correct and 
led to the prediction
\be
\lim_{L\to\infty}\frac{\Delta_L}{L} = \frac{1}{\sqrt{2}}\,\lambda^{1/4} + \cdots = \sqrt{2\,\pi\,g} + \cdots~.
\ee
The same result has also been obtained in \cite{Beccaria:2007qx} using the light-cone Bethe equations proposed in \cite{Arutyunov:2006ak}
in the general case and in \cite{Arutyunov:2005hd} for the $\mathfrak{su}(1|1)$ sector. 

These methods could not go beyond the leading term for various technical reasons. The only result at next-to-leading order is 
reported in \cite{Arutyunov:2005hd}. An analysis of the quantum string Bethe Ansatz equations and of a $\mathfrak{su}(1|1)$
truncation of the $\ads$ superstring in the uniform light-cone gauge consistently gives the quantum anomalous dimension  
\be
\lim_{L\to\infty}\frac{\Delta_L}{L} = \frac{1}{\sqrt{2}}\,\lambda^{1/4} -\frac{3}{2} + {\cal O}(\lambda^{-1/4}).
\ee
Remarkably, the next-to-leading correction exactly cancels the classical contribution to the scaling dimension.

Hence, an interesting question is that of recovering and possibly improve the above prediction starting 
from the BES integral equation describing this sector and derived in \cite{Rej:2007vm}. 
A first approach in this direction is the work \cite{Ryang:2007qr} which however does not include the dressing effects in the strong coupling analysis.
In this paper, we mimic the $\mathfrak{sl}(2)$ analysis and test the direct methods of \cite{Benna:2006nd}
and \cite{Alday:2007qf} to see if they are able to provide further information of the next-to-leading terms as well as the 
functional form of the Bethe roots distribution. In particular, we want to understand which is the origin of the 
$\Delta\sim \lambda^{1/4}$ scaling compared to the $\Delta\sim \sqrt\lambda$ regime which characterizes the $\mathfrak{sl}(2)$ sector.
We believe that this first step is mandatory if we want to apply the powerful all-order methods of \cite{Basso:2007wd} and ~\cite{Kostov:2008ax}
to this sector.

The detailed plan of the paper is the following. 
In Sec.~(\ref{sec:BES}) we recall the BES equation for the highest states in the $\mathfrak{su}(1|1)$ sector in the thermodynamical limit.
In Sec.~(\ref{sec:linear}) we derive a linear problem for the coefficient of the Neumann expansion of the solution of the BES equation.
In Sec.~(\ref{sec:NLO}) we present a NLO approximation of the exact linear problem.
Sec.~(\ref{sec:numerics}) is devoted to our numerical study of the exact and NLO linear problems. 
Finally, Sec.~(\ref{sec:shape}) is devoted to the analytical proof of the NLO expansion of the anomalous dimension at strong coupling, based on 
the numerical insight described in Sec.~(\ref{sec:numerics}). A few appendices are devoted to various technical problems and proofs.

\section{The BES integral equation for the $\mathfrak{su}(1|1)$ sector}
\label{sec:BES}

The integral equation for this sector has been derived in~\cite{Rej:2007vm} and reads
\be
\label{eq:integral-equation}
\rho(t) = e^{-t}\left\{J_0(2\,g\,t)-2\,g^2\,t\,\int_0^\infty\,dt'\left[K_{\rm m}(2\,g\,t, 2\,g\,t')+4\,K_{\rm d}(2\,g\,t, 2\,g\,t')\right]\,\rho(t')\right\},
\ee
where $\rho(t)$ is essentially the Fourier transform of the first-level Bethe roots. 
The explicit expression of the {\em main} kernel $K_{\rm m}$ is
\ba
K_{\rm m}(t, t') &=& K_0(t, t') + K_1(t, t') = \frac{J_1(t)\,J_0(t')-J_0(t)\,J_1(t')}{t-t'}, \nonumber \\
K_0(t, t') &=&  \frac{t\,J_1(t)\,J_0(t')-t'\,J_0(t)\,J_1(t')}{t^2-t^{' 2}}, \\
K_1(t, t') &=&  \frac{t'\,J_1(t)\,J_0(t')-t\,J_0(t)\,J_1(t')}{t^2-t^{' 2}}. \nonumber
\ea
The various terms admit the following Neumann expansion in series of Bessel functions
\ba
K_0(t, t') &=& \frac{2}{t\,t'}\sum_{n\ge 1}(2\,n-1)\,J_{2\,n-1}(t)\,J_{2\,n-1}(t'), \nonumber \\
K_1(t, t') &=& \frac{2}{t\,t'}\sum_{n\ge 1} 2\,n\,J_{2\,n}(t)\,J_{2\,n}(t'), \\
K_{\rm m}(t, t') &=& \frac{2}{t\,t'}\sum_{n\ge 1} n\,J_{n}(t)\,J_{n}(t').\nonumber
\ea
The {\em dressing} kernel is 
\be
K_{\rm d}(t, t') = 4\,g^2\,\int_0^\infty dt''\,K_1(t, 2\,g\,t'')\,\frac{t''}{e^{t''}-1}\,K_0(2\,g\,t'', t'),
\ee
and admits the Neumann expansion
\be
K_{\rm d}(t, t') = \sum_{n\ge 1}\sum_{m\ge 1} \frac{8\,n\,(2\,m-1)}{t\,t'}\,Z_{2\,n, 2\,m-1}\,J_{2\,n}(t)\,J_{2\,m-1}(t'),
\ee
where the coupling dependent matrix $Z_{n, m}(g)$ is
\be
Z_{n,m}(g) = \int_0^\infty\frac{dt}{t}\,\frac{J_n(2\,g\,t)\,J_m(2\,g\,t)}{e^t-1}.
\ee
Unfortunately, as far as we know, this integral cannot be computed in closed form.
Its accurate numerical evaluation is discussed in App.~(\ref{app:z}).

Given the solution of the integral equation \refeq{integral-equation}, the anomalous dimension of the $\mathfrak{su}(1|1)$ operator
${\cal O} = \mbox{Tr}\,\psi^L$ divided by the operator length can be computed in the thermodynamical limit as 
\be
\lim_{L\to \infty}\frac{\Delta}{L} = \phantom{\frac{1}{2}}2\,g^2\,E(g),
\ee
where 
\be
E(g) = 4\,\int_0^\infty dt\,\frac{J_1(2\,g\,t)}{2\,g\,t}\,\rho(t).
\ee
The full anomalous dimension is obtained by adding the classical $+3/2$ contributions, the bare
dimension of $\mbox{Tr}\,\psi^L$ divided by $L$.

\subsection{Weak coupling expansion of the integral equation}

The perturbative expansion of $\rho(t)$ for small $g$ can easily be obtained as discussed in~\cite{Rej:2007vm}. 
The initial terms are 
\ba
\rho(t) &=& e^{-t}\big\{1-t\,(1+t)\,g^2+\frac{1}{4}\,t\,(t^3+2\,t^2-2\,t+16)\,g^4 + \\
&& -\frac{1}{36}\,t\,(t^5+3\,t^4-6\,t^3+78\,t^2-180\,t+288\,\zeta_3\,t+1044)\,g^6 + \cdots\big\} ,\nonumber
\ea
and the associated energy $E(g)$ reads
\be
E(g) = -\frac{2}{g^2}(1+\rho'(0)) = 2-8\,g^2+58\,g^4-(518+32\,\zeta_3)\,g^6 + \cdots~.
\ee

\section{The equivalent linear problem}
\label{sec:linear}

We introduce the new function $\sigma(t) = e^t\,\rho(t)$ obeying the equation
\be
\sigma(t) = J_0(2\,g\,t)-2\,g^2\,t\,\int_0^\infty\,dt'\left[K_{\rm m}(2\,g\,t, 2\,g\,t')+4\,K_{\rm d}(2\,g\,t, 2\,g\,t')\right]\,e^{-t'}\,\sigma(t'),
\ee
and expand it in Neumann series
\be
\sigma(t) = \sum_{n\ge 0}J_n(2\,g\,t)\,\sigma_n,\qquad \sigma_0 = 1.
\ee
The anomalous dimension is simply
\be
\lim_{L\to\infty} \frac{\Delta}{L} = -4\,g\,\sigma_1.
\ee
The integral equation can be written
\ba
\label{eq:presystem}
\sum_{n\ge 0}J_n(2\,g\,t)\,\sigma_n &=& J_0(2\,g\,t)-\sum_{m\ge 1}\,\sum_{n\ge 0} m\,J_m(2\,g\,t)\,W_{m,n}\,\sigma_n \\
&& -16\,\sum_{m\ge 1}\sum_{k\ge 1}\sum_{n\ge 0} m\,(2\,k-1)\,J_{2\,m}(2\,g\,t)\,Z_{2\,m, 2\,k-1}\,W_{2\,k-1, n}\,\sigma_n,\nonumber
\ea
where
\be
W_{n,m}(g) = \int_0^\infty \frac{dt}{t}\,e^{-t/(2\,g)}\,J_n(t)\,J_m(t).
\ee
This integral can be evaluated in closed form with the result
\ba
\lefteqn{
W_{n,m}(g) = g^{n+m}\,\frac{\Gamma(n+m)}{\Gamma(n+1)\,\Gamma(m+1)}\times} && \\
&& \times\, {}_4 F_3\left(
\begin{array}{c}
\frac{1}{2}(n+m), \frac{1}{2}(n+m+1), \frac{1}{2}(n+m+1), \frac{1}{2}(n+m+2) \\
n+1, m+1, n+m+1
\end{array}; -16\,g^2
\right).\nonumber
\ea
We can split \refeq{presystem} into the following two equations valid for $m\ge 1$
\ba
\sigma_{2\,m} &=& -2\,m\,\sum_{n\ge 0} W_{2\,m, n}\,\sigma_n-16\,m\,\mathop{\sum_{k\ge 1}}_{n\ge 0} (2\,k-1)\,Z_{2\,m, 2\,k-1}\, W_{2\,k-1, n}\,\sigma_n, \\ 
\sigma_{2\,m-1} &=& -(2\,m-1)\,\sum_{n\ge 0} W_{2\,m-1, n}\,\sigma_n. \nonumber
\ea
The second equation can be replaced in the first obtaining our final form for the linear system equivalent to the BES integral equation

\medskip
\noindent
\underline{\em Exact linear problem}
\ba
\label{eq:exact-linear-system}
\sigma_{2\,m} &=& -2\,m\,\sum_{n\ge 0} W_{2\,m, n}\,\sigma_n+16\,m\,\sum_{k\ge 1} Z_{2\,m, 2\,k-1}\,\sigma_{2\,k-1}, \\ 
\sigma_{2\,m-1} &=& -(2\,m-1)\,\sum_{n\ge 0} W_{2\,m-1, n}\,\sigma_n. \nonumber
\ea

\subsection{Weak coupling expansion of the equivalent linear problem}

As a check we can recover from the linear problem \refeq{exact-linear-system} the weak coupling expansion of the energy. We start from 
the easy expansions
\be
Z_{n,m} = {\cal O}(g^{n+m}),\qquad W_{n,m} = {\cal O}(g^{n+m}),
\ee
finding for instance
\be
\begin{array}{lll}
Z_{2,1} &=& \zeta_3\,g^3-10\,\zeta_5\,g^5 + 105\,\zeta_7\,g^7 + \cdots, \\
Z_{2,3} &=& 2\,\zeta_5\,g^5 -35\,\zeta_7\,g^7 + \cdots, \\
Z_{2,5} &=& 3\,\zeta_7\,g^7 + \cdots, \\
\end{array}
\qquad
\begin{array}{lll}
W_{1,0} &=& g-3\,g^3+20\,g^5 + \cdots, \\
W_{1,1} &=& g^2-6\,g^4+50\,g^6 + \cdots, \\
W_{1,2} &=& g^3-10\,g^5+105\,g^7+\cdots~.
\end{array}
\ee
Then, we expand the linear system with the position 
\be
\sigma_{2\,n} = \sum_{p\ge 0} \sigma_{2\,n}^{(p)}\,g^{2\,p},\qquad
\sigma_{2\,n+1} = \sum_{p\ge 0} \sigma_{2\,n+1}^{(p)}\,g^{2\,p+1},
\ee
under the condition $\sigma_0\equiv 1$. We immediately find
\ba
\sigma_1 &=& -g+4\,g^3-29\,g^5 + \cdots, \nonumber \\
\sigma_2 &=& -g^2+(10-16\,\zeta_3)\,g^4+(-100+64\,\zeta_3+160\,\zeta_5)\,g^6 + \cdots, \\
\sigma_3 &=& -g^3+18\,g^5-240\,g^7 + \cdots, \nonumber \\
\sigma_4 &=& -g^4 + (28-32\,\zeta_5)\,g^6+(-482+128\,\zeta_5+672\,\zeta_7)\,g^8 + \cdots, \nonumber \\
&\cdots& \nonumber
\ea
which reconstructs the previous expression of $\sigma(t)$ order by order in $g^2$.

\subsection{Basso-Korchemsky zero-mode formulation}

Let us derive zero-mode equations in the spirit of \cite{Basso:2007wd}. We split 
\ba
\sigma(2\,g\,t) &=& \sigma_+(2\,g\,t) + \sigma_-(2\,g\,t), \nonumber \\
\sigma_+(2\,g\,t) &=& J_0(2\,g\,t)+\sum_{n\ge 1} J_{2\,n}(2\,g\,t)\,\sigma_{2\,n}, \\
\sigma_-(2\,g\,t) &=& \sum_{n\ge 1} J_{2\,n-1}(2\,g\,t)\,\sigma_{2\,n-1}. \nonumber
\ea
If $n$ and $m$ are positive and have the same parity, then we have 
\be
\int_0^\infty \frac{dt}{t}\,J_n(t)\,J_m(t) = \frac{1}{2\,n}\,\delta_{n,m}.
\ee
From this relation and the definitions of $W$ and $Z$ it is straightforward to derive the following equations 
($n>0$ in the first one)
\ba
\int_0^\infty \frac{d\tau}{\tau} J_{2\,n}(\tau) && \left\{
\sigma_+(\tau) + \frac{1}{2} e^{-\tau/(2\,g)}(\sigma_+(\tau)+\sigma_-(\tau))-\frac{4}{e^{\tau/(2\,g)}-1}\,\sigma_-(\tau)
\right\} = 0, \nonumber \\
\int_0^\infty \frac{d\tau}{\tau} J_{2\,n+1}(\tau) && \left\{
\sigma_- + \frac{1}{2} e^{-\tau/(2\,g)}(\sigma_+(\tau)+\sigma_-(\tau))
\right\} = 0.
\ea
Hence, 
\ba
\sigma_+(\tau) + \frac{1}{2} e^{-\tau/(2\,g)}(\sigma_+(\tau)+\sigma_-(\tau))-\frac{4}{e^{\tau/(2\,g)}-1}\,\sigma_-(\tau) &=& \chi_+(\tau), \\
\sigma_- + \frac{1}{2} e^{-\tau/(2\,g)}(\sigma_+(\tau)+\sigma_-(\tau)) &=& \chi_-(\tau), \nonumber 
\ea
where $\chi_\pm(\tau)$ are the most general solutions of the equations 
\ba
\int_0^\infty \frac{d\tau}{\tau} J_{2\,n}(\tau) && \chi_+(\tau) = 0, \qquad (n>0)\\
\int_0^\infty \frac{d\tau}{\tau} J_{2\,n+1}(\tau) && \chi_-(\tau) = 0.
\ea
As in the case of the $\mathfrak{sl}(2)$ sector, these zero-mode equations admit many solutions and 
one must find a way to fix their contributions. In this paper, we focus on the next-to-leading solution 
of the integral equation and shall not discuss in details the zero-mode equations. Notice that this is already a non-trivial task
since the non-homogeneous piece in the integral equation 
gives only $\sigma_+(0)=1$ and it can be seen that the normalization of $\sigma_-$ is not fixed by the zero-mode equations.
It must be found from the integral equation, or the equivalent linear problem.

\section{Next-to-leading order linear problem}
\label{sec:NLO}

Following the approach in~\cite{Alday:2007qf}, it will be very convenient to study an approximation to the exact linear problem \refeq{exact-linear-system}.
This is obtained by replacing the integrals $Z$ and $W$ by the first terms of their asymptotic expansions.
We choose the following level of approximation 
\ba
Z_{n,m}(g) &=& g\,Z_{n,m}^\ell -\frac{1}{2}\,W_{n,m}^\ell + \cdots, \\
W_{n,m}(g) &=& W_{n,m}^\ell + \cdots, \nonumber
\ea
where
\be
Z^\ell_{2\,m, 2\,k-1} = \frac{1}{4\,m} \left\{
\begin{array}{ll}
\displaystyle \frac{1}{2\,m-1}, & k=m \\ \\
\displaystyle \frac{1}{2\,m+1}, & k=m+1
\end{array}
\right.~,
\ee
and
\ba
W^\ell_{n,m}(g) = \int_0^\infty\frac{dt}{t} J_n(t)\,J_m(t) = \left\{
\begin{array}{ll}
\displaystyle \frac{1}{2\,n}, & n=m \\ \\
\displaystyle \frac{2}{\pi}\frac{\sin((n-m)\frac{\pi}{2})}{n^2-m^2}, & n\neq m
\end{array}
\right.~.
\ea
This approximation will be denoted as next-to-leading order (NLO). The reason is that the strong coupling expansion of $Z$ and $W$ goes in 
inverse powers of $g$ times logarithms, whereas the natural expansion parameter for the anomalous dimension is supposed to be $1/\sqrt{g}$, with possible
logarithmic corrections.

Thus, at NLO, the exact linear problem \refeq{exact-linear-system} can be written in the following form 
 ($m\ge 1$ in the first equation and $m\ge 0$ in the second one) 

\noindent
\underline{\em NLO linear problem}
\ba
\label{eq:NLO1}
\frac{3}{2}\,\sigma_{2\,m} &=& -10\,m\,\sum_{n\ge 0} W^\ell_{2\,m, 2\,n+1}\,\sigma_{2\,n+1} + 4\,g\,\left(
\frac{\sigma_{2\,m-1}}{2\,m-1}+\frac{\sigma_{2\,m+1}}{2\,m+1}\right),\\
\frac{3}{2}\,\sigma_{2\,m+1} &=& -(2\,m+1)\,\sum_{n\ge 0} W^\ell_{2\,m+1, 2\,n}\,\sigma_{2\,n}. \nonumber
\ea

Before attempting an analytical study of this problem, let us present a numerical investigation of the exact and NLO linear
problems.

\section{Numerical analysis of the linear problems}
\label{sec:numerics}

\subsection{Strong coupling expansion of the anomalous dimensions}

As a first step, we have studied numerically the exact linear problem \refeq{exact-linear-system}.
To this aim, we have truncated the mode expansion up to $N$ modes and have studied the convergence of the 
solution with increasing $N$ at various couplings $g$. Examples are shown in Fig.~(\ref{fig:exact.convergence}).

A better behavior is obtained by studying in the same way the NLO linear problem \refeq{NLO1}. This is nice as illustrated in Fig.~(\ref{fig:convergence}),
but what is the accuracy of the NLO approximation ? Reasonably, it should work well for large $g$ at least 
with the aim of reproducing the NLO expression of the strong coupling expansion. This assumption is explicitly tested in 
Fig.~(\ref{fig:Exact-NLO}). As the figure shows, the agreement between the exact and NLO calculations is quite good in the considered range. 

Given these positive results, we have attempted to extract the strong coupling expansion of the energy density by a numerical fit of the 
NLO solution at quite large values of $g$. The analysis is shown in Fig.~(\ref{fig:NLO-Fit}). The leading term $\sim g^{1/2}$ is evident.
The functional form that we have chosen
and which is rigorously suitable for the NLO problem is 
\be
\lim_{L\to\infty}\frac{\Delta}{L} = \sqrt{2\,\pi\,g}\,c_1 -\frac{3}{2}\,c_2 + \frac{\log\,g}{\sqrt{g}}\,c_3 + \frac{1}{\sqrt{g}}\,c_4 + \cdots~.
\ee
The best fit gives 
\be
|c_1-1|, \ \ |c_2-1|\sim 10^{-5}.
\ee
strongly supporting the prediction in~\cite{Arutyunov:2005hd}
\be
\lim_{L\to\infty}\frac{\Delta}{L} = \sqrt{2\,\pi\,g} -\frac{3}{2}+\cdots~.
\ee
In particular, we have recovered the intriguing cancellation of the classical $+3/2$ contribution.

The next corrections are definitely beyond the reach of the NLO approximation and must be extracted from the solution of the 
{\em exact} linear problem \refeq{exact-linear-system}, although the $\sqrt{g}$ dependence requires to push the exact calculation to quite large $g$. 
We have worked up to  $g = 180$ which is rather challenging from the computational point of view. The solution of the exact linear problem is well fitted by the following terms
\be
\label{eq:nice}
\lim_{L\to\infty}\frac{\Delta}{L} = \sqrt{2\,\pi\,g} -\frac{3}{2} + \frac{0.12(1)\,\log\,g + 0.28(1)}{\sqrt{g}} + \cdots,
\ee
as illustrated in Fig.~(\ref{fig:nnlo}). 

The agreement is rather remarkable, although we are using a simple two-parameter fit. 
Of course, the logarithmic NNLO term deserves special attention. For instance, it appears to be absent in the $\mathfrak{sl}(2)$ sector.
Therefore, despite the good numerical evidence, we prefer to be conservative and we do not exclude the possibility that it could be  a numerical fake. 
For instance, it could be due to fitting over a too narrow range of values of the coupling $g$. As another possibility, 
it could mimick a non-perturbative contribution once the log-free perturbative tail is subtracted.
In conclusion, we believe that additional numerical study at larger $g$ 
would be important to assess the NNLO corrections in \refeq{nice}
in a firm and safe way. This is a worthwhile issue in order to exclude possible order-of-limits problems in comparing 
semiclassical string theory calculations against BES-like integral equations.

\subsection{Shape analysis}

It is interesting to analyze the shape of the solution of the NLO problem at various $g$. This can be done by plotting at increasing $N$ the 
even and odd parts of $\sigma_n$ after multiplication by a trivial alternating sign, {\em i.e.}
\ba
\mbox{even part} &:& (-1)^n\,\sigma_{2\,n}, \\
\mbox{odd part} &:& (-1)^n\,\sigma_{2\,n+1}. \nonumber
\ea 
The result is shown in 
Figs.~(\ref{fig:shape.10.eps}, \ref{fig:shape.100.eps}, \ref{fig:shape.1000.eps}) for $g=10, 100, 1000$.

The conclusion is that the even/odd part of $\sigma_n$ tends to smooth curves (apart from the alternating sign)
and a non-trivial scaling shape emerges. This is an important issue that we can partially analyze quantitatively.

Indeed, from the numerics, it is natural to set
\ba
\sigma_{2\,m+1} &=& (-1)^m\,(2\,m+1)\,\frac{A}{\sqrt{g}} + \frac{\sigma_{2\,m+1}^{(1)}}{g} + \cdots, \\
\sigma_{2\,m} &=& \sigma_{2\,m}^{(0)} + {\cal O}(g^{-1/2}). \nonumber
\ea
Replacing this Ansatz into the second equation of \refeq{NLO1} and using $\sigma_0^{(0)}=1$ we find
\be
\frac{1}{(2\,m+1)^2}+\sum_{n\ge 1}\frac{(-1)^n}{(2\,m+1)^2-(2\,n)^2}\,\sigma_{2\,n}^{(0)} = 0.
\ee
A simple solution, again well supported by the numerics, is 
\be
\label{eq:even}
\sigma_{2\,m}^{(0)} = 2\,(-1)^m.
\ee
The normalization of this contribution is fixed and this is not surprising since it is related to the normalization 
$\sigma_+(0)=1$ in the zero-mode formulation. Instead, things are quite different for the odd part.

Replacing \refeq{even} into the first equation of \refeq{NLO1} we get
\be
\frac{3}{2}\cdot 2\,(-1)^m = 4\left(\frac{\sigma_{2\,m+1}^{(1)}}{2\,m+1}-\frac{\sigma_{2\,m-1}^{(1)}}{2\,m-1}\right).
\ee
The general solution is 
\be
\sigma_{2\,m+1}^{(1)} = (-1)^m\,(2\,m+1)\left(\frac{3}{8}(2\,m+1)+a\right),\qquad a, \ \mbox{constant}.
\ee
It depends on an arbitrary constant $a$ which is not fixed in this naive expansion. 
In conclusion, we have obtained the following expansion 
\ba
\sigma_{2\,m} &=& 2\,(-1)^m + \cdots, \\
\sigma_{2\,m+1} &=& (-1)^m\,(2\,m+1)\left[\frac{A}{\sqrt{g}} + \frac{1}{g}\left(\frac{3}{8}\,(2\,m+1)+a\right) + \cdots\right],\nonumber
\ea
where $A$ and $a$ are not determined. 

This indetermination is clearly related to the zero-mode contributions in the zero-mode formulation. Summing at this order
the functions $\sigma_\pm(\tau)$ ($\tau = 2\,g\,t$) we find the results 
\ba
\sigma_+(\tau) &=& \cos\,\tau + \cdots,  \\
\sigma_-(\tau) &=& \frac{1}{2}\,\tau\,J_0(\tau)\left(\frac{A}{\sqrt{g}} + \frac{a}{g}\right) +\frac{3}{16\,g}\,\tau\,\cos\,\tau + \cdots~,\nonumber
\ea
which satisfy the leading order zero mode equations.
The first term in the second equation contains the zero-mode $\chi_- = \tau\,J_0(\tau)$
with a basically undetermined coefficient plus a contribution which is inherited from the even leading zero mode $\chi_+ = \cos\tau$
whose normalization is fixed by $\sigma_+(0)=1$.

\subsection{Scaling}

We now ask ourselves the basic question whether it is possible to derive analytically the results 
\be
\label{eq:crux}
A = -\sqrt\frac{\pi}{8},\qquad a=0,
\ee
which are required to reproduce the observed strong coupling expansion of the anomalous dimension at next-to-leading order.

A simple way to achieve this result is based on the above remarks about scaling. Our expansion for the odd subsequence $\sigma_{2\,m+1}$
can be written in the following suggestive form 
\be
\label{eq:shape1}
(-1)^m\,\sigma_{2\,m+1} = 2\,A\,\frac{m+\frac{1}{2}}{\sqrt{g}} + \frac{3}{2}\,\left(\frac{m+\frac{1}{2}}{\sqrt{g}}\right)^2 + \frac{2\,a}{\sqrt{g}}\,\frac{m+\frac{1}{2}}{\sqrt{g}} + \cdots
\ee
This means that if we take the limit $g\to\infty$ with fixed $x=\frac{m+\frac{1}{2}}{\sqrt{g}}$ we find
\ba
\label{eq:shape2}
(-1)^m\,\sigma_{2\,m+1} &=& F(x) = F^{(0)}(x) + \frac{1}{\sqrt{g}}\,F^{(1)}(x) + \cdots, \nonumber \\
F^{(0)}(x) &=& 2\,A\,x + \frac{3}{2}\,x^2 + \cdots, \\
F^{(1)}(x) &=& 2\,a\,x + \cdots~. \nonumber
\ea
This result is numerically tested in Fig.~(\ref{fig:scaling.odd}).
Hence, the desired coefficients $A$ and $a$ can be extracted as
\be
A = \frac{1}{2}\,\lim_{x\to 0}\frac{F^{(0)}(x)}{2\,x},\qquad
a = \frac{1}{2}\,\lim_{x\to 0}\frac{F^{(1)}(x)}{2\,x}.
\ee
We now illustrate a way to compute the functions $F^{(0)}$ and $F^{(1)}$ with the required accuracy.

\section{The shape equation in the continuum limit}
\label{sec:shape}

Let us define for $n\ge 0$
\be
\sigma_{2\,n} = (-1)^n\,z^+_n,\qquad \sigma_{2\,n+1} = (-1)^n\,z^-_n.
\ee
The NLO linear problem can be written
\ba
\label{eq:zsystem}
\frac{3}{2}\,z_m^+ &=& -10\,m\,\sum_{n\ge 0}\frac{2}{\pi}\frac{1}{(2\,n+1)^2-(2\,m)^2}\,z_n^-+4\,g\,\left(\frac{z^-_m}{2\,m+1}-\frac{z^-_{m-1}}{2\,m-1}\right), 
\qquad m\ge 1,\nonumber \\
\frac{3}{2}\,z_m^- &=& (2\,m+1)\,\sum_{n\ge 0}\frac{2}{\pi}\frac{1}{(2\,n)^2-(2\,m+1)^2}\,z_n^+,\qquad m\ge 0.
\ea
The second equation can be inverted by using the result
\be
\frac{2}{\pi}\sum_{m\ge 0}\frac{1}{(2\,m+1)^2-(2\,p)^2}\,\frac{1}{(2\,m+1)^2-(2\,n)^2} = \left\{
\begin{array}{ll}
\displaystyle -\frac{\pi}{16\,p^2}, & n=0, \\ \\
\displaystyle \phantom{-}\frac{\pi}{32\,p^2}, & n>0,
\end{array}
\right. .
\ee
The exchange of summation is safe due to the asymptotic properties of $z_n^-$ as can be checked {\em a
posteriori} on the numerical solution. We obtain
\be
z_p^+ = 2 + \frac{48}{\pi}\,\sum_{m\ge 0}\frac{p^2}{(2\,p)^2-(2\,m+1)^2}\,\frac{z_m^-}{2\,m+1}.
\ee
Replacing in the first equation of \refeq{zsystem} and setting 
\be
\xi_n = \frac{z_n^-}{2\,n+1},
\ee
we arrive at the single equation
\be
\label{eq:alpha}
\frac{8}{3}\,g\,(\xi_p-\xi_{p-1})-\frac{48}{\pi}\sum_{m\ge 0}\frac{p^2}{(2\,p)^2-(2\,m+1)^2}\,\xi_m+\frac{40\,\alpha}{3\,\pi}\,
\sum_{m\ge 0}\frac{2\,m+1}{(2\,p)^2-(2\,m+1)^2}\,\xi_m-2=0,
\ee
where $\alpha=1$. The reason for introducing this parameter is that it will be irrelevant to the aim of computing the constant $A$. Therefore we shall 
be able to choose it in an optimal way.
At the numerical level, we can solve \refeq{alpha} by imposing as usual the condition $\xi_N=0$ for some large $N$ to be increased up to 
convergence. In Fig.~(\ref{fig:alpha}) we show the solution for four values of $\alpha$ (including the original $\alpha=1$) at $g=100$ with $N=600$ modes.
As one can see, the curve for $\xi_n$ starts at the same value. 

By the way, we immediately see that the determination of the constant $A$ in \refeq{crux} is non trivial. The problem \refeq{alpha} is solved with the boundary condition 
$\xi_n\to 0$ for $n\to \infty$. This position produces a unique solution and a non ambiguous prediction for $\xi_0$ which depends non-locally on the whole solution.

The rigorous proof of independence on $\alpha$ is illustrated in Appendix~(\ref{app:alpha}) where it is shown to hold at least in the range $0\le \alpha < 9/5$ which 
includes the default value $\alpha=1$.
Due to this remarkable feature we can make the easiest choice $\alpha=0$ and consider the simplified problem 
\be
\label{eq:discrete-shape}
\frac{8}{3}\,g\,(\xi_p-\xi_{p-1})-\frac{48}{\pi}\sum_{m\ge 0}\frac{p^2}{(2\,p)^2-(2\,m+1)^2}\,\xi_m-2=0.
\ee
We repeat that this problem is well posed in the following sense: We truncate the sequence $\{\xi_m\}$ to $0\le m \le N$ and impose
$\xi_N=0$. The above linear problem has a unique solution with a definite point-wise limit as $N\to\infty$.

\subsection{Continuum limit}

Based on the numerical analysis, we define the continuum limit of \refeq{discrete-shape} by setting $x = p/\sqrt{g}$ and assuming 
the following scaling form valid at fixed $x$ and large $g$ 
\be
\label{eq:continuum-variables}
\xi_p = \frac{1}{\sqrt{g}}\,f\left(\frac{p+1/2}{\sqrt{g}}\right) + \cdots,
\ee
where $f$ is a suitable smooth function.
The half-integer shift in the argument of $f$ is important. Indeed, the results of App.~(\ref{app:continuum}) imply that
\begin{enumerate}
\item[(i)] $f(x)$ obeys the following 
Cauchy integro-differential  {\em shape equation} 
\be
\label{eq:shape}
\begin{array}{l}
\displaystyle
f'(x)-\frac{9}{2\,\pi} \pint_0^\infty \frac{x^2}{x^2-y^2}\,f(y)\,dy -\frac{3}{4} = 0, \\
f(+\infty) = 0, 
\end{array}
\ee
The relation between $f(x)$ and $F^{(0)}(x)$ defined in \refeq{shape2} is 
\be
f(x) = \frac{F^{(0)}}{2\,x},
\ee
and therefore $A = \xi(0)$. Notice also that the shape equation immediately implies $\xi'(0) = \frac{3}{4}$ under very mild assumptions on its solution. 
This is in agreement with the second term in  $F^{(0)}(x)$. The solution of the shape equation is 
invariant under constant shifts $\xi\to \xi + \mbox{constant}$. This freedom is fixed by the condition $\xi(+\infty)=0$.
An equivalent form of the shape equation is easily obtained by manipulating the principal integral arriving at 
\be
\begin{array}{l}
\label{eq:shape-normalized}
\displaystyle
f'(x)-\frac{9}{2\,\pi} \pint_0^\infty \frac{y^2}{x^2-y^2}\,f(y)\,dy  = 0, \\ \\
\displaystyle
\int_0^\infty f(x)\,dx = -\frac{\pi}{6}, \qquad f(+\infty) = 0,
\end{array}
\ee
where now the (homogeneous) first equation has a rescaling freedom which is fixed by the area constraint in the second line.

\item[(ii)] the corrections to \refeq{continuum-variables} are ${\cal O}(1/g)$, {\em i.e.} 
\be
\xi_p = \frac{1}{\sqrt{g}}\,\left[f\left(\frac{p+1/2}{\sqrt{g}}\right) + \frac{1}{g}\,f^{(1)}\left(\frac{p+1/2}{\sqrt{g}}\right) + \cdots\right].
\ee
This immediately implies that $F^{(1)}(x)=0$ and therefore $a=0$ which is the second part of \refeq{crux}.
\end{enumerate}

To summarize, we have shown that the NLO expansion of the anomalous dimension can be proved analytically if we 
are able to show that the solution of the shape equation \refeq{shape} satisfies
\be
f(0) = -\sqrt{\frac{\pi}{8}}.
\ee
This will be proved in the next section. Later, we shall derive additional more complete information about the function $f(x)$.

\subsection{The value $f(0)\equiv A$}
\label{sec:GF}

The value $f(0)$ can be computed even without knowing the complete solution. To this aim, we write
\ba
f'(x) &=& \frac{9}{2\,\pi} \pint_0^\infty \frac{y^2}{x^2-y^2}\,f(y)\,dy = 
\frac{9}{4\,\pi} \pint_0^\infty \frac{2\,y^2+(x^2-y^2)-(x^2-y^2)}{x^2-y^2}\,f(y)\,dy = \nonumber\\
&=& \frac{9}{4\,\pi} \pint_0^\infty \frac{x^2+y^2}{x^2-y^2}\,f(y)\,dy-\frac{9}{4\,\pi}\,\int_0^\infty f(x)\,dx.
\ea
Using the known area constraint in \refeq{shape-normalized}, we find 
\be
f'(x) = \frac{9}{4\,\pi} \pint_0^\infty \frac{x^2+y^2}{x^2-y^2}\,f(y)\,dy+\frac{3}{8}.
\ee
Hence, using $f(+\infty)=0$, we get 
\ba
\label{eq:tmp1}
f(0)^2 &=& -\int_0^\infty (f(x)^2)'\,dx = -2\,\int_0^\infty f(x)\,f'(x)\,dx = \nonumber\\
&=& - \frac{9}{2\,\pi} \int_0^\infty dx\,\pint_0^\infty dy\, \frac{x^2+y^2}{x^2-y^2}\,f(x)\,f(y)-\frac{3}{4}\,\int_0^\infty f(x)\,dx.
\ea
The first term vanished by antisymmetry $x\leftrightarrow y$. The exchange of a standard integral and a principal value integral is allowed
by the results of~\cite{Tricomi}. Using again the area constraint, we find 
\be
f(0)^2 = -\frac{3}{4}\cdot \left(-\frac{\pi}{6}\right) = \frac{\pi}{8},\qquad \la\qquad f(0) = -\sqrt{\frac{\pi}{8}}.
\ee
As a comment, we remark that this result can also be obtained directly from the discrete equation \refeq{discrete-shape}. The method is analogous to the 
above derivation. We multiply by $\xi_p+\xi_{p-1}$ and take a sum over $p$. After symmetrization of the double sum, one easily proves the above boundary value
as the truncation number $N\to\infty$.

\subsection{Complete solution of the shape equation} 
\label{sec:solution}

A canonical form is obtained by scaling $z = \frac{3}{\sqrt{2}}\,x$ and setting $f\left(\frac{\sqrt{2}}{3}\,z\right) \equiv h(z)$. It reads
\be
\begin{array}{l}
\displaystyle
h'(z)-\frac{1}{\pi} \pint_0^\infty \frac{w^2}{z^2-w^2}\,h(w)\,dw  = 0, \\ \\
\displaystyle
\int_0^\infty h(z)\,dz = -\frac{\pi}{2\,\sqrt{2}}, \qquad h(+\infty) = 0.
\end{array}
\ee
To solve the shape equation in canonical form we follow the methods of~\cite{Spence}. We  
change variable and define $h(\sqrt{u}) \equiv g(u)$. In the new variable $u$, the equation reduces to 
\be
g'(u)-\frac{1}{4\,\pi}\,\pint_0^\infty \frac{\sqrt{t}}{1-t}\,g(u\,t)\,dt.
\ee
The Mellin transform~\cite{Friot:2005gh}
\be
\widetilde{g}(s) = \int_0^\infty u^{s-1}\,g(u)\,du,
\ee
has fundamental strip $(0,1)$ and obeys the difference equation 
\be
\label{eq:difference}
\frac{\widetilde{g}(s)}{\widetilde{g}(s-1)} = -4\,(s-1)\,\cot\,\pi\,s.
\ee
The solution which is associated with a $g(u)$ regular in $u=0$ and vanishing for $u\to\infty$ is easily shown to be 
\be
\widetilde{g}(s) = 4^s\,\Gamma(s)\,\Phi(-s)\,\kappa,
\ee
where $\kappa$ is a constant to be determined by the area constraint (or equivalently by the condition on $f'(0)$).
The function $\Phi(s)$ is the ratio
\be
\Phi(s) = \frac{G(1-s)\,G(\frac{3}{2}+s)}{G(1+s)\,G(\frac{1}{2}-s)},
\ee
where $G(z)$ is the Barnes $G$-function discussed for instance in~\cite{Barnes} and defined as 
\be
G(z+1) = (2\,\pi)^{z/2}\,\exp\left(-\frac{1}{2}\,z\,(z+1)-\frac{1}{2}\,\gamma_{\rm E}\,z^2\right)\,\prod_{n\ge 1}\left(1+\frac{z}{n}\right)^n\,
e^{-z+\frac{z^2}{2\,n}}.
\ee
It obeys in particular
\ba
G(z+1) &=& \Gamma(z)\,G(z), \\
G(1) &=& 1, \nonumber
\ea
from which it is easy to prove that the difference equation is satisfied. Useful properties of the ratio $\Phi(s)$ are the infinite product expression 
\be
\Phi(s) = \sqrt{\pi}\,\prod_{n=1}^\infty\frac{\left(1-\frac{s}{n}\right)^n\,\left(1+\frac{s}{n+\frac{1}{2}}\right)^n}
{\left(1+\frac{s}{n}\right)^n\,\left(1-\frac{s}{n-\frac{1}{2}}\right)^n},
\ee
the functional relations
\be
\Phi(s) = \Phi(-s-\frac{1}{2}) = \Phi(s-1)\,\tan\,\pi\,s, 
\ee
and the expansion in $s=0$
\be
\Phi(s) = \sqrt{\pi}\left[1+s+\left(1+\frac{\pi^2}{2}\right)\,\frac{s^2}{2!}+\left(1+\frac{17\,\pi^2}{6}\right)\,\frac{s^3}{3!} + \cdots\right].
\ee
From these relations we can look at the poles for negative half-integer $s$ and for positive integer $s$ and use the known properties of the Mellin transform 
to derive asymptotic expansions of $g(u)$ around $u=0$ and $u=+\infty$. The result is 
\ba
\kappa^{-1}\,g(u) &=& \sqrt{\pi}-u^{1/2}+u^{3/2}\left(-\frac{1}{6\,\pi}\,\log\,u+\frac{11-3\,\gamma_E}{18}\right) + {\cal O}(u^{5/2}\,\log^2\,u), \\
\kappa^{-1}\,g(u) &=& \frac{4}{\sqrt{\pi}}\frac{1}{u}+\frac{16}{\pi^{3/2}}\,\frac{1}{u^2}(-\log\,u+2\,\log\,2-\gamma_E) + {\cal O}(u^{-3}\,\log^2\,u). \nonumber
\ea
Taking $\kappa = -\sqrt{2}/4$ we recover the correct expansion
\be
f(x) = -\sqrt\frac{\pi}{8}+\frac{3}{4}\,x + \cdots.
\ee
As an additional  check of the solution, we can evaluate the area constraint finding again agreement
\ba
\int_0^\infty h(z)\,dz &=& \frac{1}{2}\widetilde{g}\left(\frac{1}{2}\right) = -\frac{\sqrt{2}}{8}\lim_{s\to 1/2} 4^s\,\Gamma(s)\,\Phi(-s) = \\
&=& 
-\frac{\sqrt{2}}{8}\cdot 2\,\cdot\Gamma(1/2)\,\sqrt{\pi} = -\frac{\pi}{2\,\sqrt{2}}\nonumber.
\ea
A numerical comparison of the above asymptotic expansions with the numerical solution of the 
shape equation is illustrated in Fig.~(\ref{fig:asymptotic}).

\section{Conclusions}

In summary, we have considered the BES-like equation describing the anomalous dimension of highest states $\mbox{Tr}\,\psi^L$ in the 
$\mathfrak{su}(1|1)$ sector for $L\to\infty$ proposed in~\cite{Rej:2007vm}. We have followed the approach pioneered in \cite{Benna:2006nd,Alday:2007qf}
for the $\mathfrak{sl}(2)$ twist operators. It is based on the Neumann expansion of the Fourier transformed Bethe root density.
Although many technical differences arise, we have been able to analytically derive the next-to-leading order
expansion of the anomalous dimension at strong coupling. Some interesting information on the NNLO terms is also presented based on numerical calculations only.
Our results reproduce at leading order the result already discussed in the literature, although with different methods. At next-to-leading order, 
we find agreement with the proposal in \cite{Arutyunov:2005hd}. Besides, several specific interesting features appear. In particular, at NLO
we have found and solved a singular integral equation describing the scaling continuum limit of the Neumann coefficients.

\acknowledgments

We thank G. F. De Angelis for important comments on the shape equation. We also thank 
S. Frolov, G.~E.~Arutyunov, A.~Tseytlin, I.~Swanson, and V.~Forini for useful discussions and helpful comments.

\appendix
\section{Numerical evaluation of $Z_{n,m}(g)$}
\label{app:z}

A simple strategy to numerically evaluate $Z_{n,m}(g)$ with arbitrary precision is based on the exact splitting 
\be
Z_{n,m}(g) = \sum_{n=0}^N\int_0^\infty \frac{dt}{t} e^{-(n+1)\,t/(2\,g)} J_n(t)\,J_m(t) + \int_0^\infty \frac{dt}{t} \frac{e^{-(N+1)\,t/(2\,g)}}{e^{t/(2\,g)}-1}\, J_n(t)\,J_m(t).
\ee
The first finite sum can be computed in terms of the known function  $W_{n,m}(g)$ since 
\be
\int_0^\infty \frac{dt}{t} e^{-(n+1)\,t/(2\,g)} J_n(t)\,J_m(t) = W_{n,m}\left(\frac{g}{n+1}\right).
\ee
The second integral can be computed by splitting the integration interval in subintervals separated by the roots of the product
$J_n(t)\,J_m(t)$. The resulting sub-integrals build an alternating sum whose terms can be computed numerically without any difficulty.
The optimal value of $N$ is $N\sim g$, but of course the result is independent on $N$.

\section{On the $\alpha$-independence}
\label{app:alpha}

The shape equation for $\alpha=0$ is \refeq{shape-normalized} that we repeat for the reader's convenience
\be
f'(x)-\frac{9}{2\,\pi} \pint_0^\infty \frac{y^2}{x^2-y^2}\,f(y)\,dy  = 0.
\ee
If the parameter $\alpha$ is not zero, the same steps starting from \refeq{alpha} give a similar result with the simple
substitution in the numerator of the integral
\be
y^2\la y^2-\beta\,x\,y,\qquad \beta = \frac{5}{9}\,\alpha.
\ee
We now show that for $0\le\beta < 1$ the first two terms in the asymptotic expansion of $f(x)$ around $x=0$ do not depend on $\alpha$.

To this aim and with the notation of Sec.~(\ref{sec:solution}) we immediately find that \refeq{difference} is replaced by 
\be
\frac{\widetilde{g}(s)}{\widetilde{g}(s-1)} = -4\,(s-1)\,\frac{1}{\tan\,\pi\,s+\beta\,\cot\,\pi\,s}.
\ee
Setting 
\be
\widetilde{g}(s) = \kappa\,4^s\,\Gamma(s)\,\widetilde{g}_0(s),
\ee
we find 
\be
\frac{\widetilde{g}_0(s)}{\widetilde{g}_0(s-1)} = -\frac{1}{\tan\,\pi\,s+\beta\,\cot\,\pi\,s}.
\ee
From now on, let us consider the case $0\le\beta < 1$. The zeroes of the denominator of the right hand side are located at the positions
\ba
s_n &=& n\pm i\,a,\qquad n\in\mathbb{Z}, \\
a &=& \frac{1}{\pi}\mbox{arctanh}\sqrt\beta. \nonumber
\ea
Looking at the zeroes and poles (with their order) of a $\widetilde{g}_0$ which is regular in the fundamental strip $0\le s< 1$ we find the solution 
\be
\widetilde{g}_0(s) = \frac{G(1-s)}{G(1+s)}\,\frac{G(\frac{3}{2}-s)}{G(\frac{1}{2}+s)}
\,\frac{G(1+i\,a+s)}{G(1+i\,a-s)}\,\frac{G(1-i\,a+s)}{G(1-i\,a-s)}\,(\cosh\, \pi\,a)^{2\,s}.
\ee
where $G(s)$ is the Barnes $G$-function introduced in Sec.~(\ref{sec:solution}). The analytic structure is manifest in the Weierstrass infinite product form 
which is 
\be
\widetilde{g}_0(s) = \sqrt{\pi}\,(\cosh\, \pi\,a)^{2\,s}\,\prod_{n\ge 1}\left[
\frac{1-\frac{s}{n}}{1+\frac{s}{n}}\,
\frac{1-\frac{s}{n+\frac{1}{2}}}{1+\frac{s}{n-\frac{1}{2}}}\,
\frac{1+\frac{s}{n+i\,a}}{1-\frac{s}{n-i\,a}}\,
\frac{1+\frac{s}{n-i\,a}}{1-\frac{s}{n+i\,a}}
\right]^n
\ee
The area constraint and the asymptotic expansion of $\widetilde{g}(s)$ have a $a$-dependence which is all contained in the factor
\be
p(s, a) = \frac{G(1+i\,a+s)}{G(1+i\,a-s)}\,\frac{G(1-i\,a+s)}{G(1-i\,a-s)}\,(\cosh\, \pi\,a)^{2\,s}.
\ee
The area constraint on the total integral $\int_0^\infty f(x)\,dx$ is $a$-independent since
\ba
p\left(\frac{1}{2}, a\right) &=& \frac{G(\frac{3}{2}+i\,a)}{G(\frac{1}{2}+i\,a)}\,\frac{G(\frac{3}{2}-i\,a)}{G(\frac{1}{2}-i\,a)}\,\cosh\, \pi\,a = \nonumber\\
&=& \Gamma(\frac{1}{2}+i\,a)\,\Gamma(\frac{1}{2}-i\,a)\,\cosh\,\pi\,a = \pi.
\ea
The coefficients of the leading and subleading terms of $f(x)$ for small $x$ are also $a$-independent since they are 
proportional to 
\be
p(0, a) = 1, \qquad p\left(-\frac{1}{2}, a\right) = \frac{1}{\pi}.
\ee
Finally, the next correction does depend on $a$ since
\be
p(-1, a) = \frac{1}{\pi^2}\,\tanh^2(\pi\,a) = \frac{\beta}{\pi}.
\ee
As a final remark, we notice that the $\alpha$-independence of $f(0)$ can be proved easily using the method
of Sec.~(\ref{sec:GF}), assuming always $f(+\infty)=0$. Indeed the $\alpha$-term just add a $x\,y$
contribution to the numerator in the double integral appearing in \refeq{tmp1}. Since this is symmetric under
$x\leftrightarrow y$, the proof valid for $\alpha=0$ is completely unchanged.

\section{Detailed continuum limit of the shape equation}
\label{app:continuum}

Let us consider \refeq{discrete-shape} with the position $p=x\,\sqrt{g}$ and $\xi_p = \frac{1}{\sqrt{g}}\,f\left(\frac{p+1/2}{\sqrt{g}}\right)$.
We shall analyze the large $g$ expansion at fixed $x$ for the two terms of  \refeq{discrete-shape} to next-to-leading accuracy.
The finite difference is immediately expanded and gives
\be
g\,(\xi_p-\xi_{p-1}) = f'(x) + \frac{1}{24\,g}\,f'''(x) + {\cal O}(1/g^2).
\ee
The infinite sum is more complicated. We use the Euler-MacLaurin formula
\be
\sum_{m=a}^b F(m) = \int_a^b F(m)\,dm + \frac{F(a)+F(b)}{2} + \sum_{k\ge 1}\frac{B_{2\,k}}{(2\,k)!}\left[
F^{(2\,k-1)}(b)-F^{(2\,k-1)}(a)
\right],
\ee
where
\be
F(m) = \frac{(2\,p)^2}{(2\,p)^2-(2\,m+1)^2}\,\frac{1}{\sqrt{g}}\,f\left(\frac{m+\frac{1}{2}}{\sqrt{g}}\right).
\ee
It is convenient to split the infinite sum in order to easily display the appearance of a principal value at leading order.
Thus, we write
\be
S = \sum_{m\ge 0} F(m) = \left(\sum_{m=0}^{p-1}+\sum_{m=p}^\infty\right)\,F(m).
\ee
Applying the Euler-MacLaurin formula we get
\be
\label{eq:S}
S =  \left(\int_{0}^{p-1}+\int_{p}^\infty\right)\,F(m)\,dm + \frac{F(0)+F(p-1)+F(p)}{2} + R_p,
\ee
where $R_p$ is the sum involving higher odd derivatives of $F$.
The two integrals can be written
\ba
\lefteqn{\left(\int_{0}^{p-1}+\int_{p}^\infty\right)\,F(m)\,dm = \left(
\int_{\frac{1}{2\,\sqrt{g}}}^{x-\frac{1}{2\,\sqrt{g}}}+
\int_{x+\frac{1}{2\,\sqrt{g}}}^\infty
\right)\,\frac{x^2}{x^2-y^2}\,f(y)\,dy = } && \\
&& =  \pint_0^\infty \frac{x^2}{x^2-y^2}\,f(y)\,dy + \frac{-2\,f(0)-f(x)+2\,x\,f'(x)}{4\,\sqrt{g}}+{\cal O}\left(\frac{1}{g}\right). \nonumber
\ea
The other two terms in the right hand side of \refeq{S} have the expansion
\ba
\frac{F(0)+F(p-1)+F(p)}{2} &=& \frac{2\,f(0)+f(x)-2\,x\,f'(x)}{4\,\sqrt{g}}+{\cal O}\left(\frac{1}{g}\right), \\
R_p &=& {\cal O}\left(\frac{1}{g}\right).
\ea
In conclusion, summing the various terms, we find 
\be
S = \sum_{m\ge 0} F(m) = \pint_0^\infty \frac{x^2}{x^2-y^2}\,f(y)\,dy +{\cal O}\left(\frac{1}{g}\right),
\ee
with complete cancellation of the $1/\sqrt{g}$ correction.
 
\newpage
\FIGURE{
\vspace*{0.6cm}
\epsfig{file=Figures/exact.convergence.eps,height=12cm}
\vspace*{0.5cm}
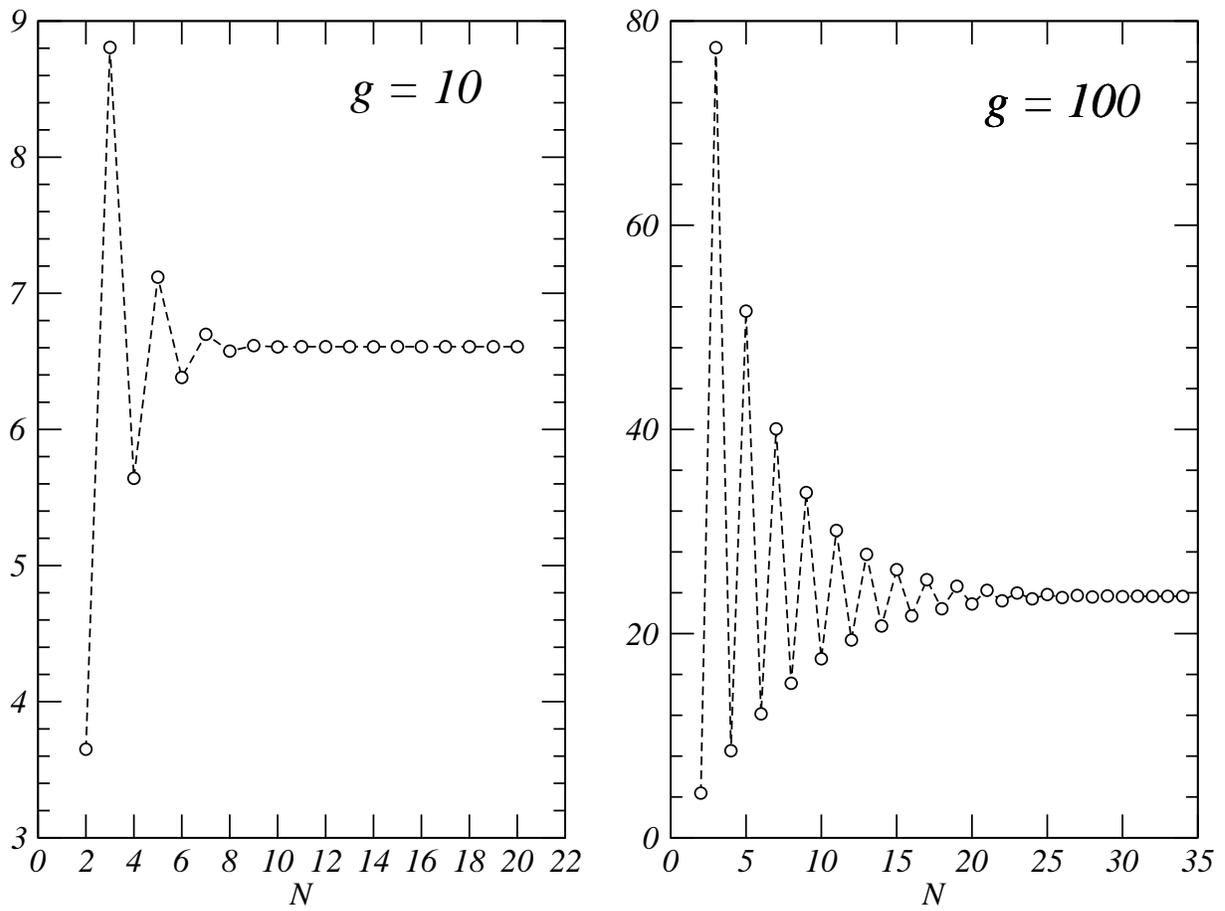
\caption{Convergence of the solution $\sigma_1$ of the exact linear problem \refeq{exact-linear-system} as the number
of truncated modes $N$ is increased. The convergence is clearly slowing down as the coupling increases.}
\label{fig:exact.convergence}
}

\newpage
\FIGURE{
\vspace*{0.6cm}
\epsfig{file=Figures/convergence.eps,height=12cm}
\vspace*{0.5cm}
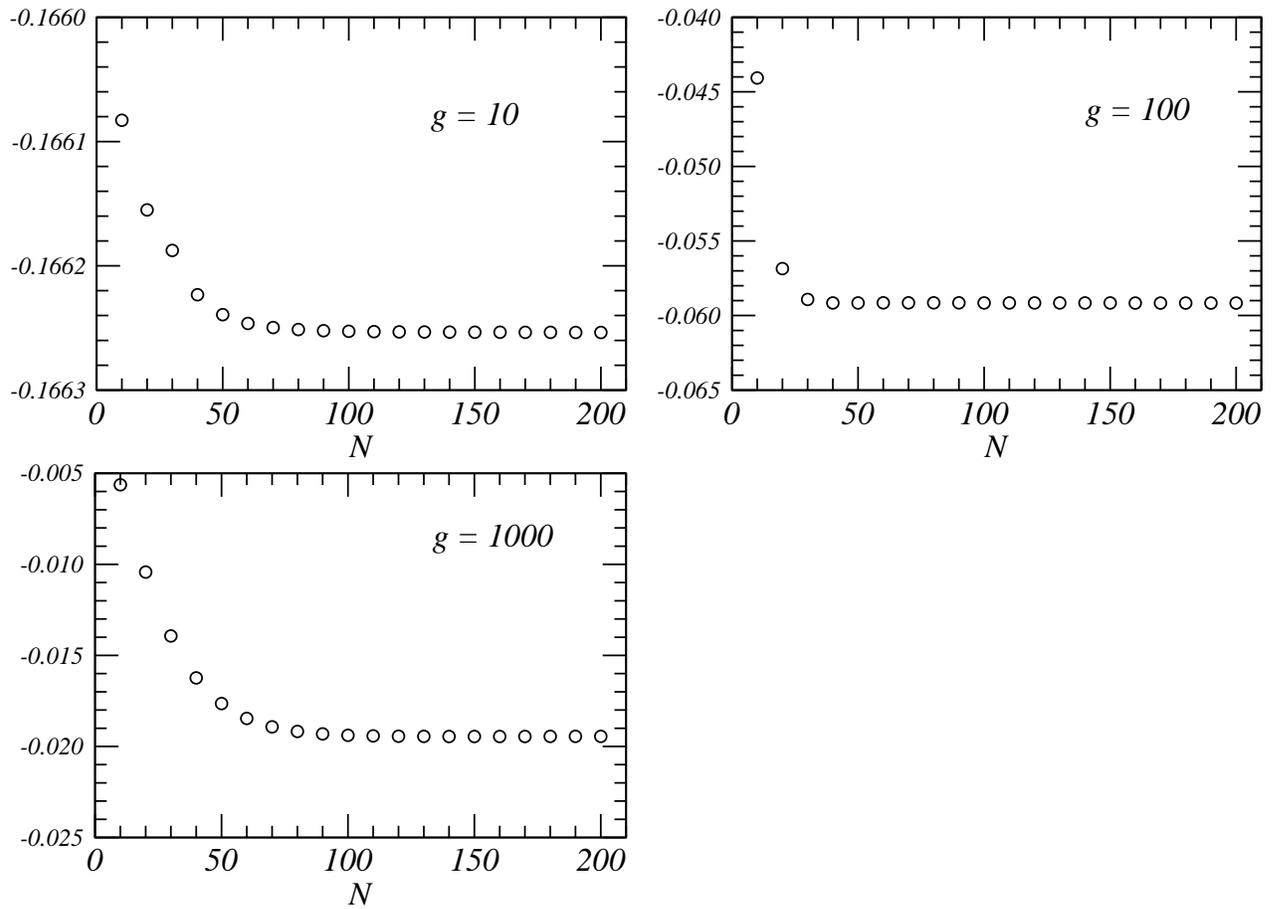
\caption{Convergence of the solution of the NLO linear problem \refeq{NLO1} as the number
of truncated modes $N$ is increased. The convergence slowing down is not is real problem even at rather strong values of the 
coupling.}
\label{fig:convergence}
}

\newpage
\FIGURE{
\vspace*{0.6cm}
\epsfig{file=Figures/Exact-NLO.eps,height=12cm}
\vspace*{0.5cm}
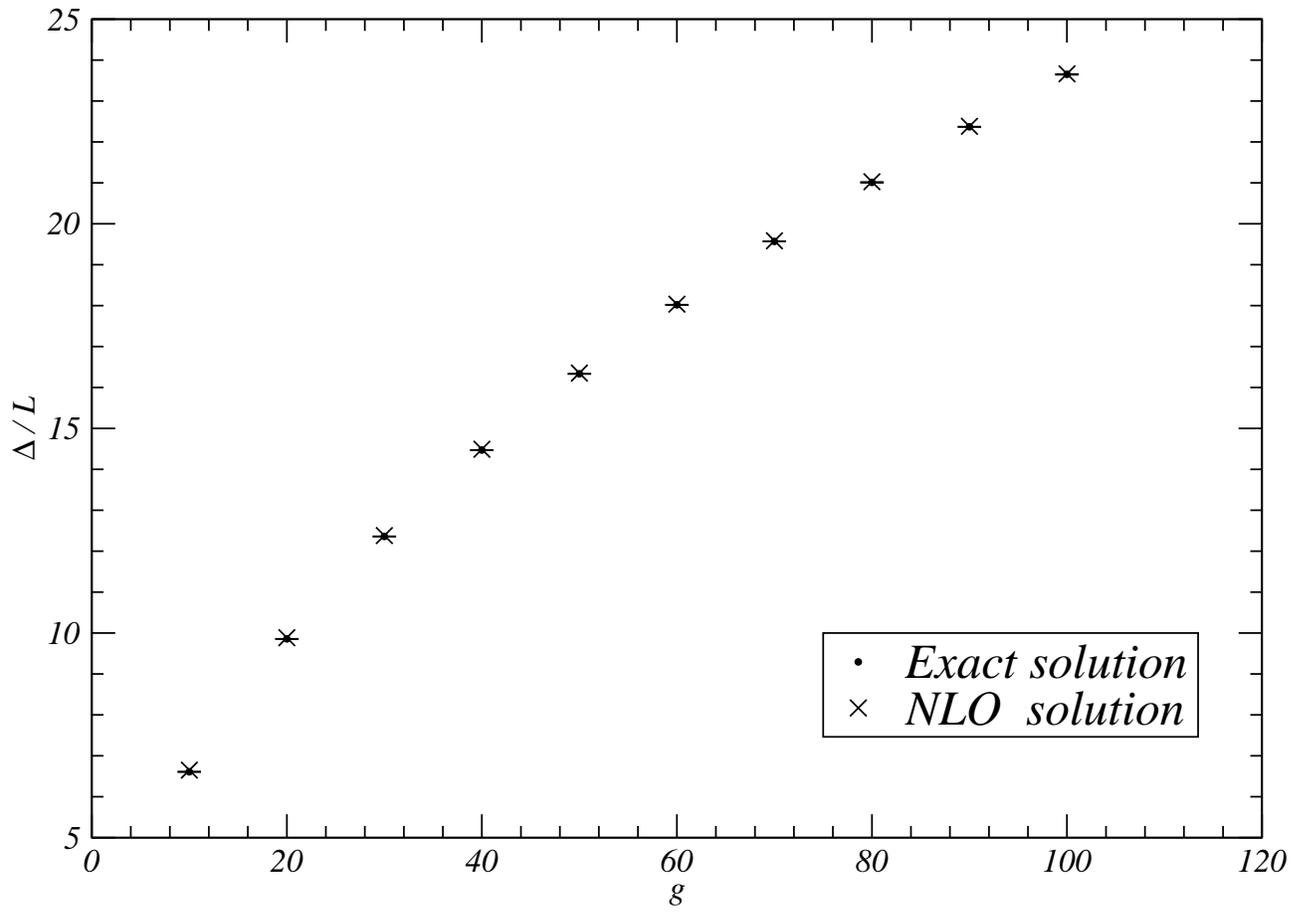
\caption{Comparison between the solutions of the exact linear problem \refeq{exact-linear-system} and its NLO approximation \refeq{NLO1}.
For all values of the coupling, the number of  truncated modes $N$ has been increased up to convergence.}
\label{fig:Exact-NLO}
}

\newpage
\FIGURE{
\vspace*{0.6cm}
\epsfig{file=Figures/NLO-Fit.eps,height=12cm}
\vspace*{0.5cm}
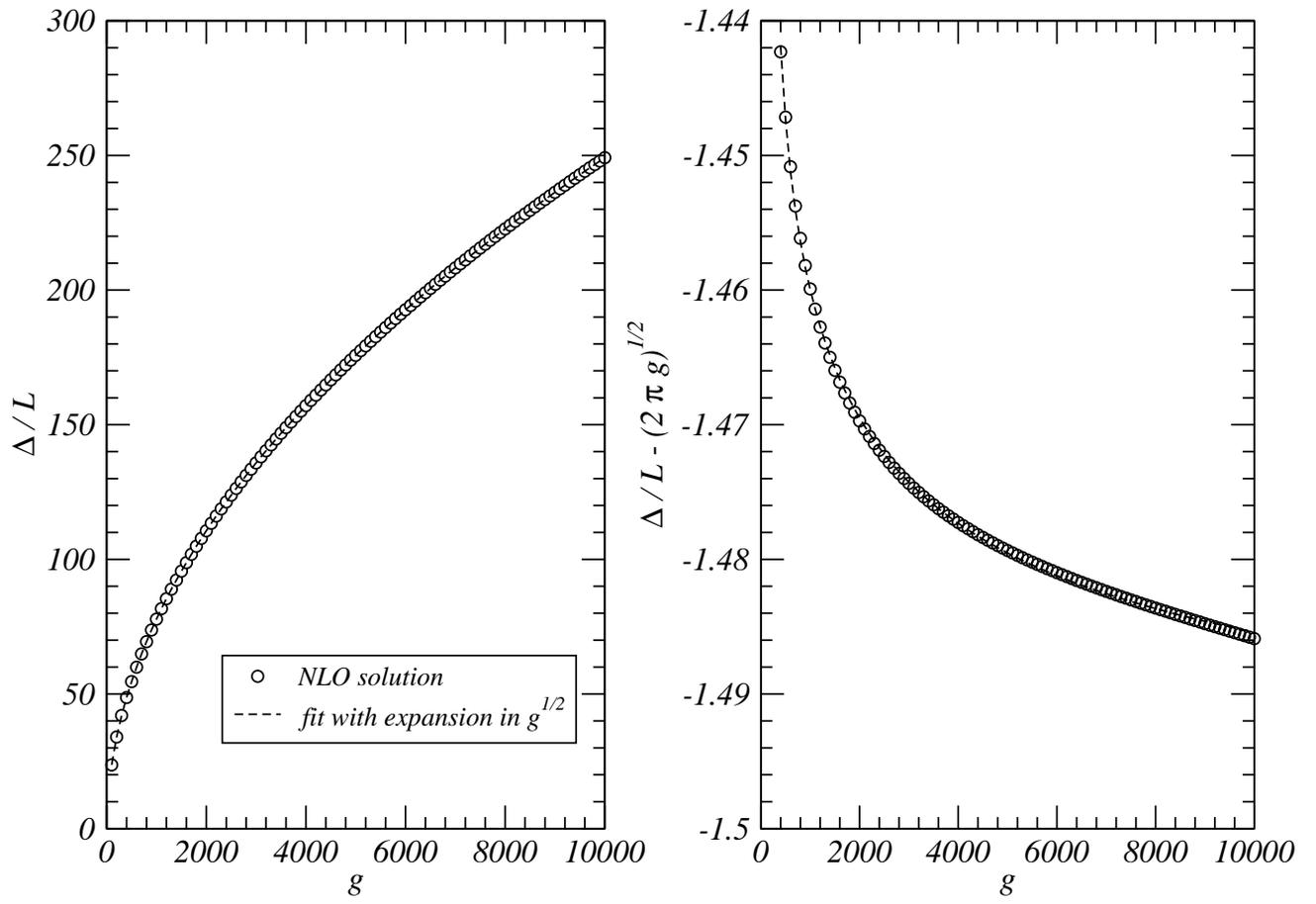
\caption{Analysis of the leading and subleading strong coupling expansion of the energy density from the numerical solution of the 
NLO linear problem \refeq{NLO1}.}
\label{fig:NLO-Fit}
}

\newpage
\FIGURE{
\vspace*{0.6cm}
\epsfig{file=Figures/NNLO.eps,height=12cm}
\vspace*{0.7cm}
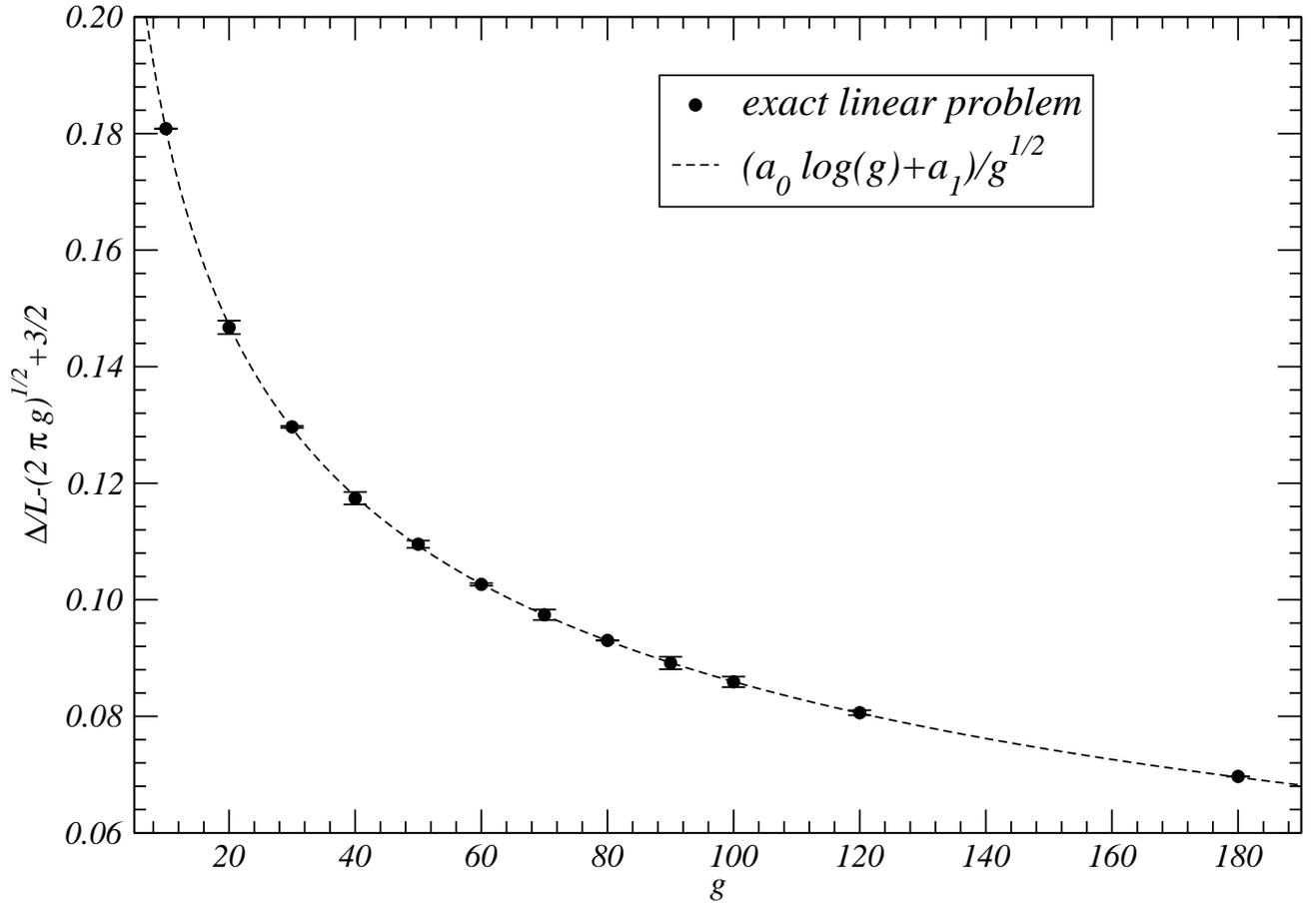
\caption{Numerical fit of the NNLO correction to the strong coupling expansion of the anomalous dimension. The best fit values of the two 
parameters are $a_0 = 0.12(1)$ and $a_1 = 0.28(1)$. These numbers are basically unchanged if the first data points (at smaller $g$) are removed from the fit.
Also, the point with the largest $g=180$ is very well reproduced by the best fit computed with the points with $g<180$. This is a check of the proposed functional form.
}
\label{fig:nnlo}
}

\newpage
\FIGURE{
\vspace*{0.6cm}
\epsfig{file=Figures/shape.g_10.eps,height=12cm}
\vspace*{0.7cm}
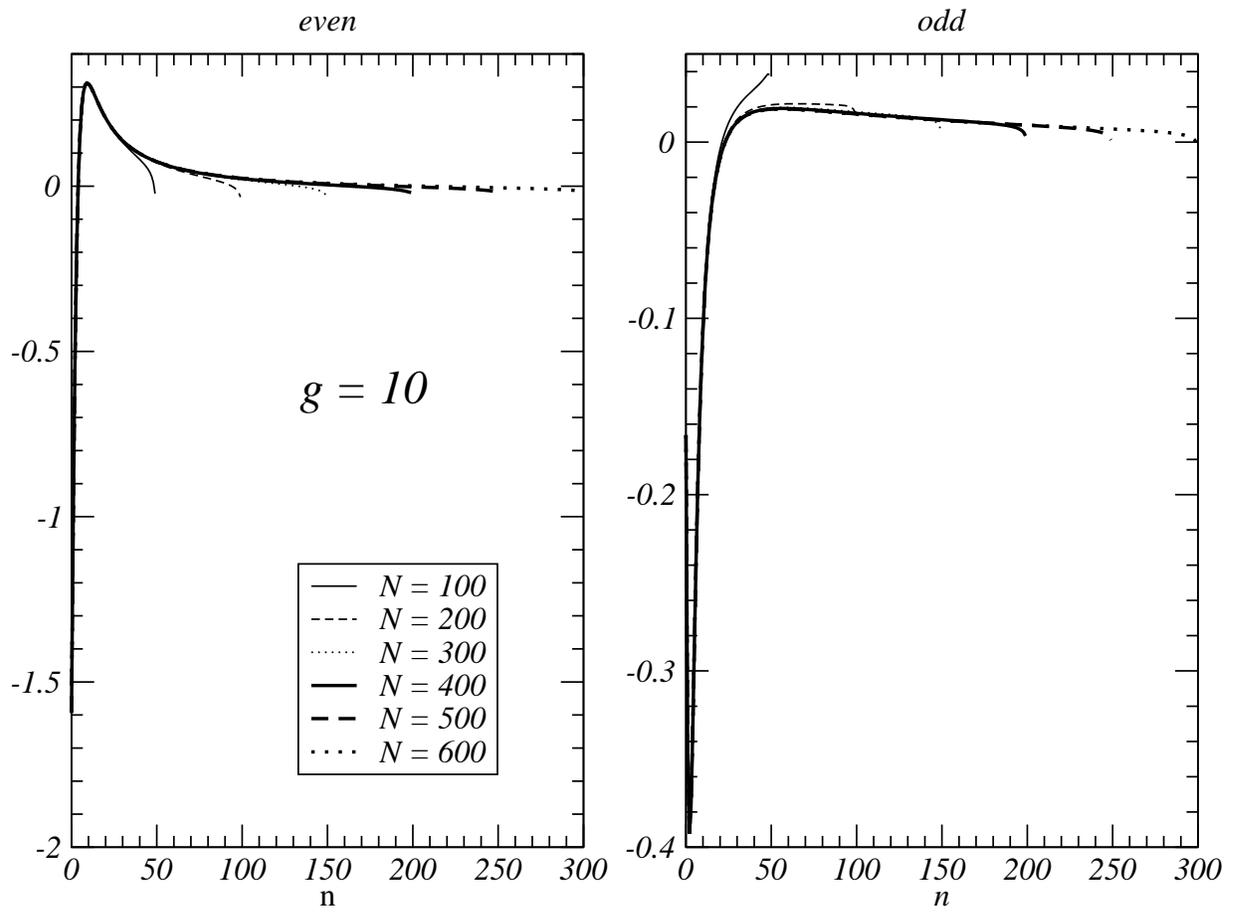
\caption{Shape of the NLO solution at $g=10$ as the mode cut $N$ is increased.}
\label{fig:shape.10.eps}
}

\newpage
\FIGURE{
\vspace*{0.6cm}
\epsfig{file=Figures/shape.g_100.eps,height=12cm}
\vspace*{0.7cm}
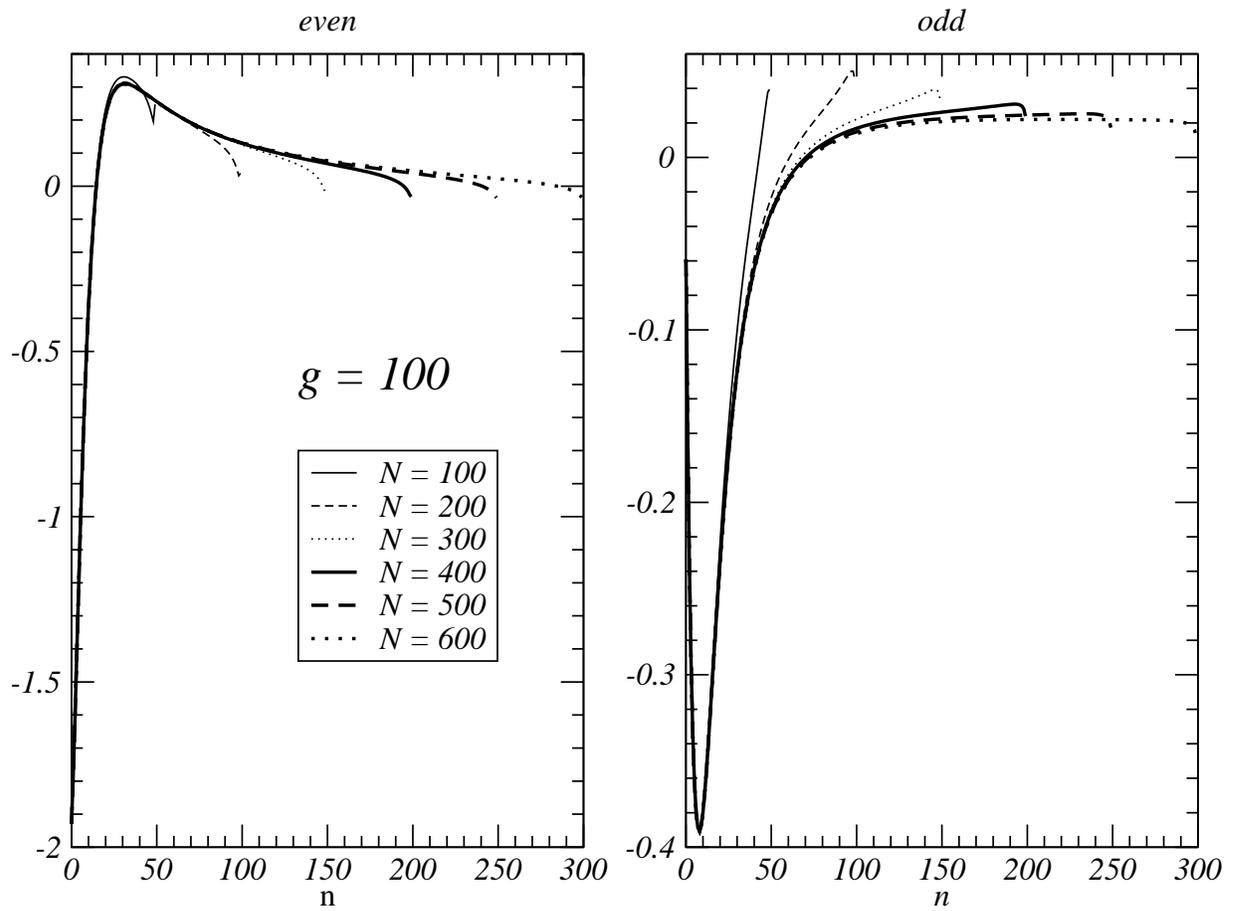
\caption{Shape of the NLO solution at $g=100$ as the mode cut $N$ is increased.}
\label{fig:shape.100.eps}
}

\newpage
\FIGURE{
\vspace*{0.6cm}
\epsfig{file=Figures/shape.g_1000.eps,height=12cm}
\vspace*{0.7cm}
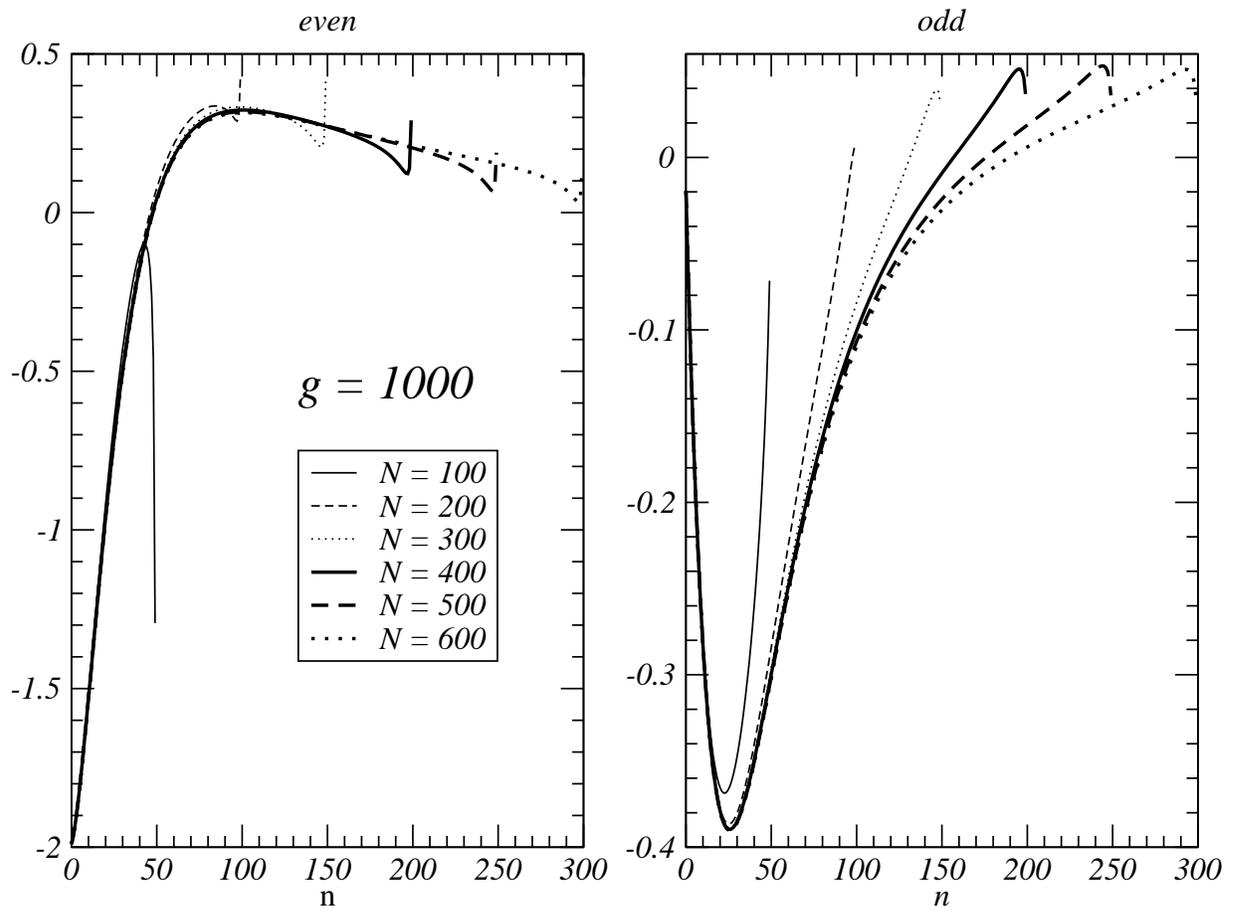
\caption{Shape of the NLO solution at $g=1000$ as the mode cut $N$ is increased.}
\label{fig:shape.1000.eps}
}

\newpage
\FIGURE{
\vspace*{0.6cm}
\epsfig{file=Figures/scaling.odd.eps,height=12cm}
\vspace*{0.7cm}
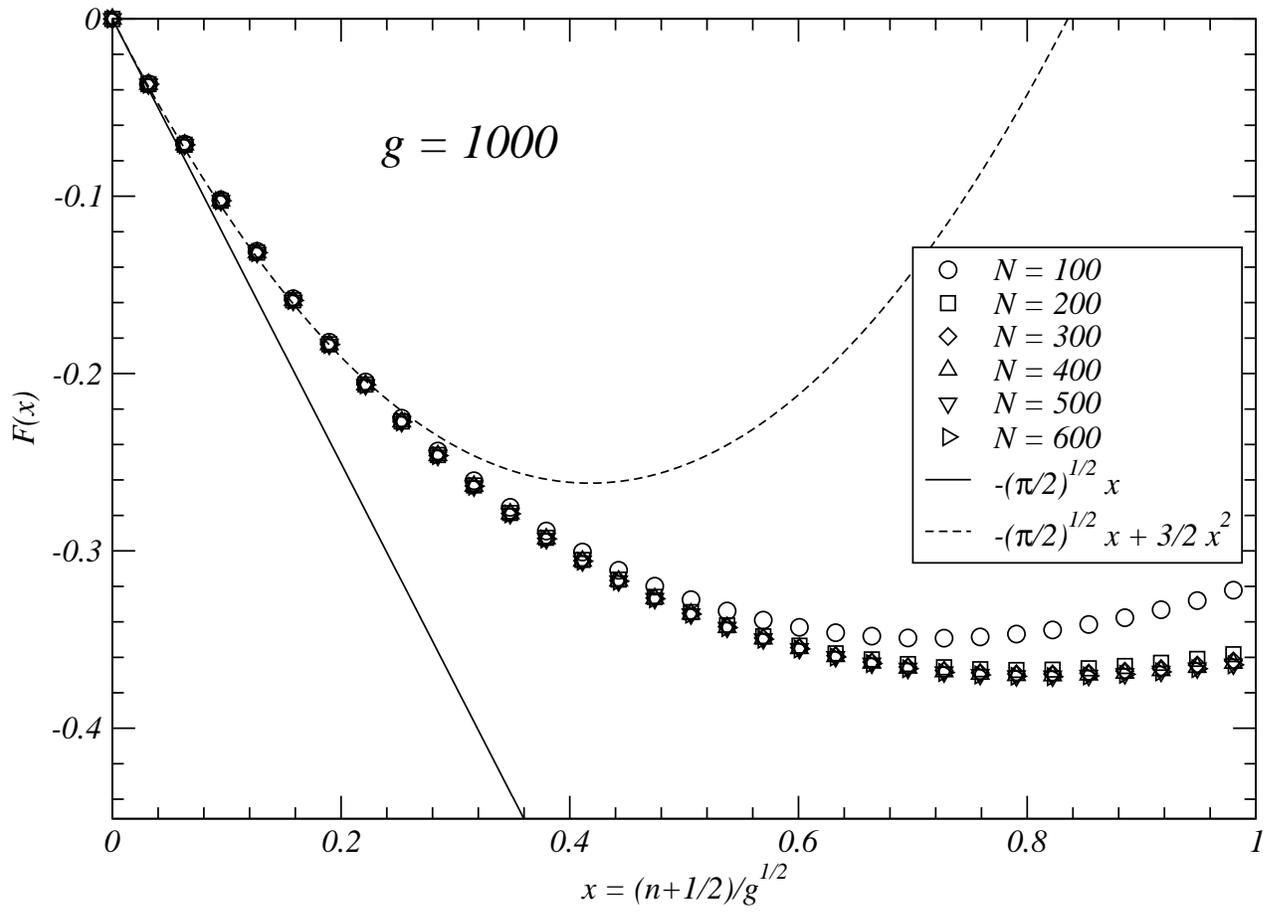
\caption{Scaling analysis. Odd part of the NLO solution at $g=1000$. The numerical shape is compared with the 
analytical result \refeq{shape1} with the correct choice \refeq{crux}.}
\label{fig:scaling.odd}
}

\newpage
\FIGURE{
\vspace*{0.6cm}
\epsfig{file=Figures/alpha.eps,height=12cm}
\vspace*{0.7cm}
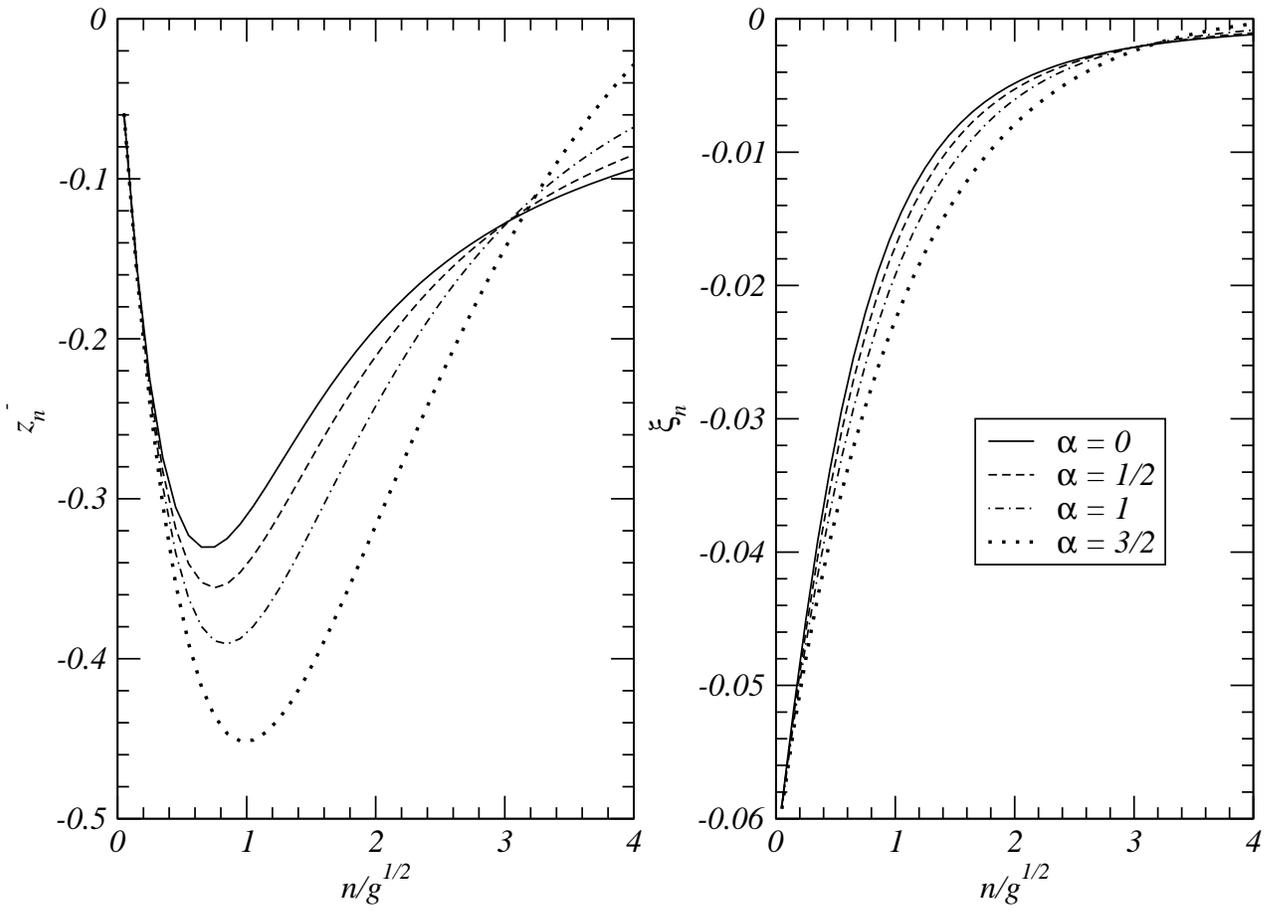
\caption{Independence of $\xi_0$ on the parameter $\alpha$ appearing in \refeq{alpha}.
The four curves are all obtained with $g=100$ and $N=600$.}
\label{fig:alpha}
}

\newpage
\FIGURE{
\vspace*{0.6cm}
\epsfig{file=Figures/asymptotic.eps,height=12cm}
\vspace*{0.7cm}
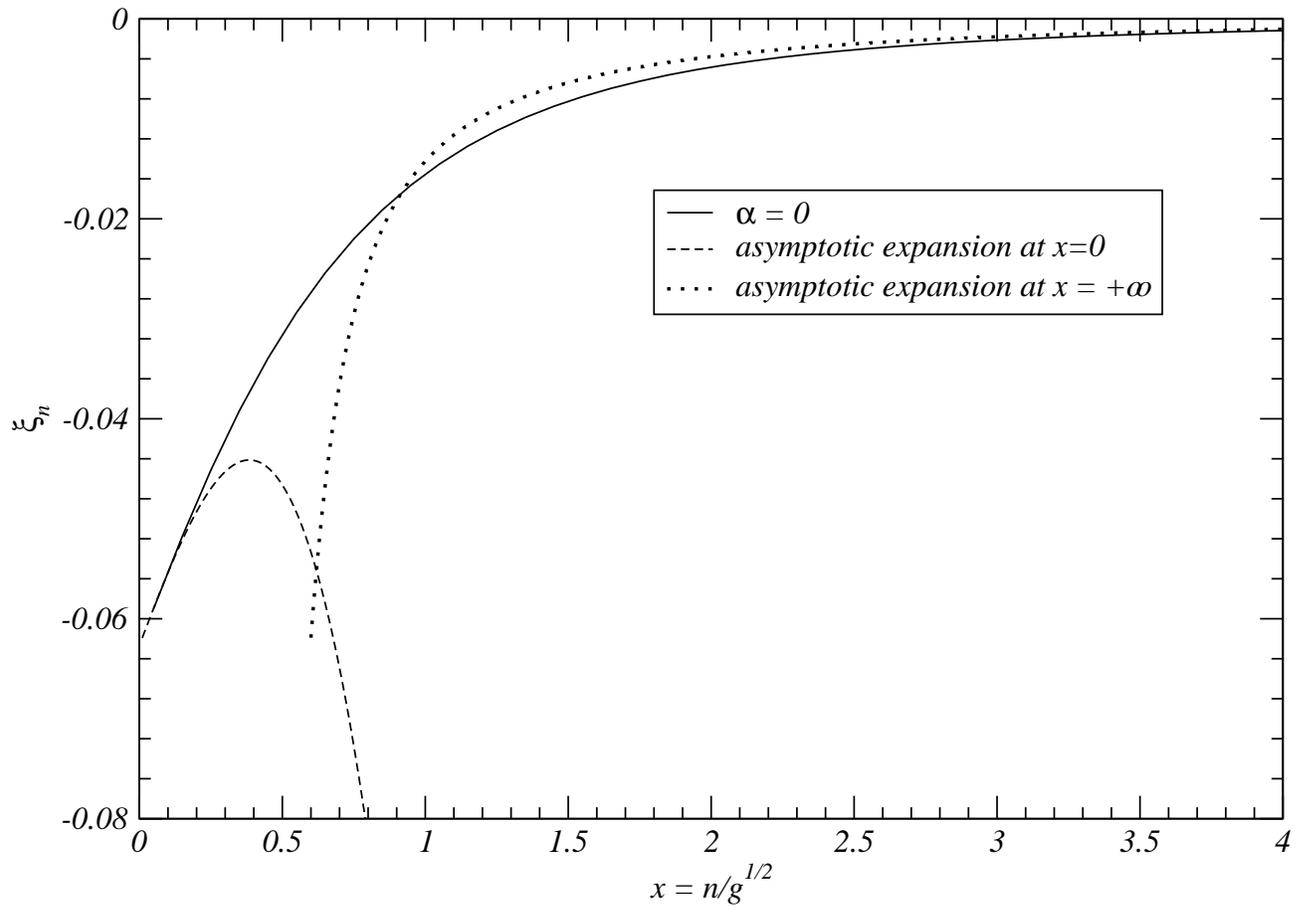
\caption{Comparison between the solution of the NLO linear problem at $\alpha=0$ and the asymptotic expansion of 
the shape equation at $x=0$ and $+\infty$. 
The first curve is obtained with $g=100$ and $N=600$.}
\label{fig:asymptotic}
}

\end{document}